\documentclass[%
onecolumn, 12pt, 
 amsmath,amssymb,
 aps,
]{revtex4-2}
\usepackage[english]{babel}
\usepackage[utf8]{inputenc}
\usepackage{dcolumn}
\usepackage{soul}
\newcommand\T{\rule{0pt}{2.5ex}}       
\newcommand\B{\rule[-1.ex]{0pt}{0pt}} 
\usepackage{graphicx}
\usepackage{bm,amsmath,amssymb}
\usepackage[mathscr]{eucal}
\usepackage{makeidx}
\usepackage{subfig}
\usepackage{amsmath}
\usepackage{amssymb}
\usepackage{wasysym}
\usepackage{amsthm}
\usepackage{mathrsfs}
\usepackage{graphicx}
\usepackage{fancyhdr}
\usepackage{array}
\usepackage{simplewick}
\usepackage{latexsym}
\usepackage[all]{xy}
\usepackage{enumerate}
\usepackage{dsfont}
\usepackage{slashed}
\usepackage{float}
\usepackage[titletoc]{appendix}
\usepackage{ulem}
\usepackage{xcolor}
\usepackage{hyperref}

\begin{document}

\title{Bosonic and Fermionic Holographic Fluctuation and Dissipation at finite temperature and density}

\author{Nathan G. Caldeira}
\email{nathangomesc@hotmail.com}
\author{Carlos A. D. Zarro}
\email{carlos.zarro@if.ufrj.br} 
\author{Henrique Boschi-Filho} 
\email{boschi@if.ufrj.br}
\affiliation{%
 Instituto de F\'{\i}sica,
Universidade Federal do Rio de Janeiro, RJ
21941-909 -- Brazil
}%


\begin{abstract}
In this paper we investigate some general aspects of fluctuation and dissipation in the holographic scenario at zero and finite density. We model this situation with a probe string in a diagonal metric representing a black brane. The string stretches  from the black brane to a probe brane thus simulating a stochastic driven particle. In this scenario, we compute the admittance, the diffusion coefficient, the correlation functions and the regularized mean square displacement, for bosons and fermions, all from the metric components. We check these calculations with the fluctuation-dissipation theorem. Further, 
we show that at  finite temperature and density, the mean square displacement in the limit of short times reproduces the usual quadratic (ballistic) behavior, for bosons and fermions. For large times,  we find  ultraslow diffusive processes in various cases, except for bosons at zero chemical potential. 
We apply this general analysis in two different models: hyperscaling violation at finite temperature  and a charged dilatonic AdS black hole, both for bosons and fermions. This is important because we found the fermionic diffusion in systems which allow the appearance of Fermi surfaces and Fermi liquids.
\end{abstract}

\maketitle
	
\tableofcontents

\section{Introduction}

The study of the dynamics of finite temperature and/or finite chemical potential systems occupies an important part in physical sciences.
Of particular interest is how such systems respond to an external force and its diffusive process. Huge results in this direction can be made studying the linear regime of those processes and the related quantities obtained from this analysis. Thus it is
interesting to explore methods to get relevant quantities in this context.

As a tool for this study we can use the framework of AdS/CFT correspondence in its broader form, applying its dictionary, for example, to analyze and describe aspects
of this kind of phenomenon.

The study of linear response in the context of Brownian motion using holographic models started in \cite{deBoer:2008gu} where it was proposed a set up based on an stretched string that goes from the black hole horizon to a probe brane near the space-time boundary. The presence of a horizon associated with a black hole allows one to calculate the Hawking temperature and to obtain some quantities as functions of temperature. One then interpret the string endpoint as a probe particle in a thermal bath. From the excitation of the string at the boundary one reads the motion of the particle and finds the diffusion coefficient from the admittance. They also calculate the mean square displacement, for short and long times reproducing the ballistic and diffusive regimes, respectively. 

This set up was used to study quantum critical points with Lifshitz symmetry in \cite{Tong:2012nf} and with hyperscaling-Lifshitz symmetry in  \cite{Edalati:2012tc}. In \cite{Giataganas:2018ekx} a description in terms of a metric written as monomials of the holographic coordinate was presented and applied to many different cases encompassing the previous results \cite{deBoer:2008gu, Tong:2012nf, Edalati:2012tc} and discussing further cases. 
In particular, in Ref. \cite{Edalati:2012tc} a  chemical potential at zero temperature was introduced in the case of the extremal black branes. This was also discussed in \cite{Banerjee:2015vmo} where a very low temperature compared to the chemical potential was introduced. 

In Ref. \cite{Caldeira:2020sot} an exponential factor was used to deform the AdS-Schwarzschild metric. This deformation was inspired by the soft-wall model where a quadratic exponential dilaton is included in the action \cite{Karch:2006pv}. This exponential factor was introduced in the metric in Ref. \cite{Andreev:2006ct} to guarantee confinement in a quark-antiquark potential. This metric was used in Ref. \cite{Caldeira:2020sot} 
to obtain the admittance, the diffusion coefficient, compute the regularized mean square displacement and verify the fluctuation-dissipation theorem for this set up. In Ref. \cite{Caldeira:2020rir}, this problem was reanalyzed including a backreaction from the exponential factor altering the horizon function, generalizing the results of Ref. \cite{Caldeira:2020sot}. 
In Ref. \cite{Caldeira:2021izy}, this set up with an exponential deformation in the metric was further extended to include a non-zero chemical potential allowing the discussion of bosonic and fermionic mean square displacements and also to verify the fluctuation-dissipation theorem in both (bosonic and fermionic) cases. In particular, the long time  regime for both cases present a Sinai-like diffusion logarithm behavior 
\cite{sinai-like}. 

In this work we extend some finite temperature  results \cite{deBoer:2008gu, Tong:2012nf, Edalati:2012tc,Giataganas:2018ekx} regarding the linear response and diffusion in the literature to a general diagonal background metric. It is important to mention that this previous results are restricted to bosons. 
We include a non-zero chemical potential in this general metric allowing us to distinguish between the bosonic and fermionic cases. 
Using the corresponding statistical distributions we calculate and present the values of the mean square displacement for bosonic and fermionic cases. After obtaining closed expressions, we get the limits of short and long times for zero and non-zero chemical potentials for bosons and fermions. This is important because the diffusion and mean square displacement were not discussed previously in the fermionic case, as we present here (we discussed the fermionic case for an exponential deformed metric which is a particular case of the present discussion). 

As an application for these general results we apply our findings on two holographic models of interest. First, we consider the hyperscaling-Lifshitz model at finite temperature and zero chemical potential for bosons and fermions. This background was studied as a holographic dual for a series of condensed matter systems and it was used in the holographic study of Fermi surfaces \cite{Huijse:2011ef}. Second, we discuss a charged dilatonic AdS black hole presented in Ref.  \cite{Gubser:2009qt}, as another application of our general results.  This is a top-down model with finite chemical potential which describes dual Fermi liquids with massless charged fermionic modes. In this model, the  entropy is a linear function of the temperature, for low $T$. So, our general results allow us to obtain the diffusion coefficient and mean square displacement of fermionic systems which are relevant to the study of Fermi surfaces and Fermi liquid in the holographic set up.

This work is organized as follows. In Section \ref{NG}, we start setting the class of metrics explored here and calculating the Hawking temperature. From the Nambu-Goto action we find the general equation of motion in the linear regime. Then we solve it using a patching method. In Section \ref{sec:Admittance}, we find a general expression for the linear response function and for the diffusion coefficient in terms of the general metric elements. 
In Section \ref{sec:corrfunct}, we calculate correlation functions in the bosonic and fermionic cases. 
In Section \ref{sec:flutuation-dissipation}, 
we check our results for the admittance and correlation functions with the fluctuation-dissipation theorem for bosons and fermions. 
In section \ref{sec:msd}, our goal is to calculate the regularized mean square displacement $s^{2}_{\text{reg}}(t)$ for multiple scenarios. We begin establishing the boundary conditions for the solutions and from them we get a general integral representation for $s^{2}_{\text{reg}}(t)$. After that, we specialize to four  different cases: bosonic and fermionic at zero and non-zero chemical potentials. For each of these cases we further obtain the regimes of short and large times. At the end of this section we present a summary of this results and compare them. 
Going ahead, in section \ref{sec:hyp}, we apply the previous results to a hyperscaling violation  $d+2$ spacetime metric at finite temperature and zero chemical potential presenting  some interesting cases for particular values of the parameters of the model. 
In section \ref{sec:GRM}, 
 we look at a top-down finite chemical potential model in $2+1$ dimensions introduced in \cite{Gubser:2009qt}. We  use it as an illustration for our general finite chemical results in the bosonic and fermionic cases presenting  interesting properties. 
Finally in the last Section \ref{sec:conc}, we show our conclusions making general comments about the results. Some technical calculations are presented in three appendices.

\section{Nambu-Goto action and equations of motion}\label{NG}

In this section we study the dynamics of the probe string, finding its equations of motion and their solutions. The set up we are going to use \cite{deBoer:2008gu} considers a probe string in bulk with a black hole. One endpoint of the string interacts with the horizon (IR) and the other to a brane near the boundary (UV). The endpoint near the UV behaves as a particle  simulating a stochastic motion.  This set up was used to describe Lifshitz  as well as hyperscaling violation \cite{Tong:2012nf,Edalati:2012tc,Giataganas:2018ekx}.

We start with a general diagonal metric
\begin{equation} 
    ds^{2}=-g_{tt}dt^{2}+g_{rr}dr^{2}+\sum_{i=1}^{d} g_{ii}dx_i^{2},
    \label{gen_metric}
\end{equation}
where we defined
\begin{equation}\label{eq:metriccomponents}
    g_{tt}\equiv a(r)f(r) \,;
    \qquad  g_{rr}\equiv\frac{b(r)}{f(r)}\,;
     \qquad  g_{ii}\equiv c(r)\,;\quad (i=1,\cdots, d).
\end{equation}
Note that $d$ is the number of spatial  dimensions and $r$ is the holographic coordinate. The horizon function $f(r)$ of the black hole is assumed to have  a simple zero at  $r=r_{h}$ and goes to 1 for $r\to \infty$. 

In the particular case of asymptotically AdS spaces,  $a(r)$, $c(r)$ and $1/b(r)$ go to $r^2$ as $r\to \infty$, where the boundary is located.  
  {On the other hand, in the case of hyperscaling violation, discussed in Sec. \ref{sec:hyp}, the metric coefficients are given by
\begin{equation}\label{eq:hypabc}
     a(r)=r^{2(z-\theta/d)} \,;
    \qquad  b(r)=r^{-2(1+\theta/d)} \,;
     \qquad   c(r)=r^{2(1-\theta/d)}\,,  
\end{equation}
which also reduce to the AdS case when $z\to 1$ and $\theta\to 0$.}

The Hawking temperature for the general metric, Eq. 
\eqref{gen_metric}, is given by 
\begin{equation}
\label{general_Hawking}
    T=\frac{1}{4\pi}\sqrt{\frac{a(r_{h})}{b(r_{h})}[f'(r_{h})]^{2}}
    =\frac{|f'(r_{h})|}{4\pi}\sqrt{\frac{a(r_{h})}{b(r_{h})}}.
\end{equation}

Let us now describe the motion of the probe string using the Nambu-Goto action:
\begin{equation}\label{ng}
S_{NG} 
=
-\frac{1}{2 \pi \alpha'} \int d\tau d\sigma \sqrt{-\gamma}\, ,
\end{equation}
\noindent where $\alpha'$ is the string tension, $\gamma = {\rm det} (\gamma_{\alpha \beta})$ and $\gamma_{\alpha \beta} = g_{mn} \partial_{\alpha}X^m \partial_{\beta}X^n $ is the induced metric on the worldsheet with $m,n = 0,1,\cdots, d+1$. 

We choose the usual static gauge, where $t = \tau$, $r = \sigma$ and $X^m= X^m(\tau, \sigma)$.  By using the general metric Eq.\eqref{gen_metric} and expanding the Nambu-Goto action keeping up to quadratic terms $\dot{X}^2$, $X'^2$, one gets: 
\begin{equation}\label{ngapprox}
S_{NG} \approx - \frac{1}{4 \pi \alpha'} \int d\tau d\sigma \left[ \; \frac{\dot{X}^2 \,g_{ii}(r) \sqrt{g_{tt}(r) g_{{rr}}(r)}}{g_{tt}(r)}-\frac{X'^2 \,g_{ii}(r) \sqrt{g_{tt}(r) g_{{rr}}(r)}}{g_{{rr}}(r)}\; \right]\,,
\end{equation}
\noindent where $\dot{X}=\partial_{\tau =t} X$ and $X'=\partial_{\sigma =r} X$. 
From this action one obtains the EoM and define $g_{xx}=g_{ii}$ (without sum over $i$ from now on) for any $i$, 
\begin{eqnarray}
&&\frac{\partial }{\partial r}\left(\frac{X'(t,r) \left(g_{xx}(r) \sqrt{g_{tt}(r) g_{{rr}}(r)}\right)}{g_{{rr}}(r)}\right) + \frac{\ddot{X}(t,r) \left(g_{xx}(r) \sqrt{g_{{tt}}(r) g_{{rr}}(r)}\right)}{g_{{tt}}(r)}=0\,.\qquad 
\end{eqnarray}
Then, by using the  decomposition $X(t,r)=e^{i\omega t}h_{\omega}(r)$, the EoM reads:
\begin{eqnarray}
&&\frac{\partial }{\partial r}
\left( h_{\omega}'(r) \frac{
g_{xx}(r) \sqrt{g_{{tt}}(r)}
}{
\sqrt{g_{{rr}(r)}}}
\right) 
+
\frac{\omega ^2 g_{xx}(r) \sqrt{ g_{rr}(r)}}{
\sqrt{g_{tt}(r)}}
h_{\omega}(r)=0\,. 
\label{equation_motion}
\end{eqnarray}
For a general background metric this equation cannot be solved analytically. Then, we will apply a standard patching method to obtain approximate analytical solutions, to be presented in the next section.

\subsection{General solution in the hydrodynamic limit}\label{sec:Hydro}

In this section we are going to consider approximate solutions for Eq. \eqref{equation_motion}  considering three different particular cases, following Refs.\cite{deBoer:2008gu,  Tong:2012nf, Edalati:2012tc, Giataganas:2018ekx}:

A) Near the horizon (IR);

B) Hydrodynamic limit: $\omega\to 0$; 

C) Far from the horizon (UV).  
\par\noindent 
Then, we will match these expressions to get an approximate solution for the Eq. \eqref{equation_motion} near UV, but keeping essential data from the IR and the hydrodynamic limit. This is useful to describe the motion of the string endpoint simulating the stochastic behavior. 

First, we consider the solution for region A:
\begin{eqnarray}
\label{hA}
    h_{A\omega}(r)&=&\frac{  {A_{1}(\omega)}}{\sqrt{g_{xx}(r_{h})}}\left[1-i\frac{\omega}{f'(r_{h})}\sqrt{\frac{b(r_{h})}{a(r_{h})}}\log\left(\frac{r}{r_{h}}-1\right)\right]\cr
    &=&\frac{  {A_{1}(\omega)}}{\sqrt{g_{xx}(r_{h})}}
    \left[1- i\frac{\omega}{4\pi T} \log\left(\frac{r}{r_{h}}-1
    \right)\right]
\end{eqnarray}
where $  {A_{1}(\omega)}$ is a normalization factor. This solution is obtained in Appendix \ref{sec:appA}. 

Now we solve the equation of motion for the hydrodynamic limit which is a general solution in $r$ but only up to order $\omega$. In this case one can neglect the term proportional to $\omega^{2}$ in Eq. \eqref{equation_motion} 
\begin{equation}
\label{reg2}
    \frac{d}{d r}\left(\frac{g_{xx}\sqrt{g_{tt}}}{\sqrt{g_{rr}}}\frac{dh_{\omega}}{dr}\right)=0,
\end{equation}
  {which implies that we are considering the condition
\begin{eqnarray}
\omega ^2  \ll \frac{\sqrt{g_{tt}(r)}}{g_{xx}(r) \sqrt{ g_{rr}(r)}}
=
\frac{\sqrt{a(r)}f(r)}{\sqrt{ b(r)} c(r) }\,.
\label{cond}
\end{eqnarray}
In general, close to the boundary we can take the metric coefficients as monomials 
$$
a(r)\sim r^{a_{t}}\, ,\, b(r)\sim r^{a_{r}}\, 
,
\,c(r)\sim r^{a_{x}}\, .
$$
Using the condition in Eq. \eqref{cond} in the UV ($r\to r_{b}$) one gets the constraint 
\begin{eqnarray}
\label{condUV}
 a_{t}-a_{r}-2a_{x}\geq 0\,, 
\end{eqnarray}
which is saturated by the asymptotic AdS case with $a(r)$, $c(r)$ and $1/b(r)$ going to $r^2$ as $r\to \infty$. Note that in the IR, $r \to r_{h}+\delta$, where $\delta$ is a small finite positive quantity ($\delta>0$)  usually called the ``stretched horizon'' (see for instance 
 \cite{deBoer:2008gu} and references therein), the condition  Eq. \eqref{cond} implies
\begin{eqnarray}
\omega ^2  \ll 
\frac{4\pi T}{g_{xx}(r_{h})}(r-r_{h})=\frac{4\pi T}{g_{xx}(r_{h})}\delta\,.
\label{condIR}
\end{eqnarray}
}
This is the hydrodynamic limit considered from now on in this work.

Solving Eq. \eqref{reg2}, we obtain
\begin{eqnarray}
\label{first_order_solution}
    h^{B}_{\omega}(r)
    &=&
    B_{1}(\omega) 
    \int^{r}\frac{\sqrt{g_{rr}}}{g_{xx}
    \sqrt{g_{tt}}}dr'
    +B_{2}(\omega)
    \cr 
    &=& B_{1}(\omega) 
    \int^{r}\frac{\sqrt{b(r')}}{g_{xx}(r')f(r')\sqrt{a(r')}}dr'
    +B_{2}(\omega)\,, 
\end{eqnarray}
where $  {B_{1}(\omega)}$ and $  {B_{2}(\omega)}$ are integration constants with respect to $r$ but are functions of $\omega$. This can be seen as the general solution up to order $\omega^{2}$ of the equation of motion.
 
One can approximate the   integral above in the IR using the fact that $f(r)$ has a simple zero at the horizon. Then we can write
\begin{eqnarray}
    h_{\omega \rm (IR)}^{B} &\approx&\frac{\sqrt{b(r_{h})}B_{1}(\omega)}{g_{xx}(r_{h})\sqrt{a(r_{h  })}}
    \int^{r}\frac{1}{f(r')}dr'+B_{2}(\omega)
    \cr 
    &=&\frac{B_{1}(\omega)}{
    4\pi T g_{xx}(r_{h}) }
    \log\left(\frac{r}{r_{h}}-1\right)+B_{2}(\omega)\,. 
\end{eqnarray}
In order to obtain the functions $  {B_{1}(\omega)}$ and $  {B_{2}(\omega)}$, we compare this expression with the solution Eq. \eqref{hA} for the deep IR region:
\begin{equation}
 B_{1}(\omega)=-i   {A_{1}(\omega)} \omega\sqrt{g_{xx}(r_{h})},\quad
    B_{2}(\omega)=\frac{  {A_{1}(\omega)}}{\sqrt{g_{xx}(r_{h})}}\,, 
\end{equation}
where $  {A_{1}(\omega)}$ can be determined by a convenient normalization. 

For the case far from the horizon (case C: UV region) we  use the solution for case B  \eqref{first_order_solution}  
\begin{eqnarray}
\label{h_B_UV}
h_{\omega}^{C} (r) \approx 
h^{B}_{\omega(\rm UV)}(r)&=&B_{1}(\omega)
\int^{r}_{r_h+\epsilon}\frac{\sqrt{b(r')}}{g_{xx}(r')f(r')\sqrt{a(r')}}dr'
+B_{2}(\omega)
\cr
&=&\frac{  {A_{1}(\omega)}}{
\sqrt{g_{xx}(r_{h})}}
\left(1
-i\omega g_{xx}(r_{h})
\int^{r}_{r_h+\epsilon}
\frac{
\sqrt{b(r')}}{
g_{xx}(r')f(r')
\sqrt{a(r')}}dr'\right).
\end{eqnarray}
In the last expression  we are interested mainly in the asymptotic behavior of $h_{\omega}^{C} (r)$ for large~$r$. 

  {This solution will be used to describe the motion of the string close to the boundary brane $r \to r_{b}$. Please note that this solution is valid in the regime of small frequencies, in particular, it can be seen as an expansion for $\omega/T \ll 1$. This becomes clear when we notice that the integral in Eq.  \eqref{h_B_UV} is dominated by its near horizon part ($r=r_{h}$) since  $f(r)$ has a simple zero in this region ($f(r) \approx f'(r_{h})(r-r_{h})$), while  $a(r)$ and $b(r)$ are regular functions at $r=r_h$. Using these facts one can approximate Eq. \eqref{h_B_UV} by
\begin{eqnarray}
\label{log}
h^{B}_{\omega(UV)}(r \sim r_{b})
&\approx&
\frac{A_{1}(\omega)}{\sqrt{g_{xx}(r_{h})}}\left[1
-i \frac{\omega\sqrt{b(r_{h})}}{f'(r_{h})
\sqrt{a(r_{h})}}
\log\left(\frac{1}{\epsilon}\right)
-i\, \Xi 
-i\omega\, \Omega(r_b) 
\right]
\cr
&=&
\frac{A_{1}(\omega)}{\sqrt{g_{xx}(r_{h})}}\left[1
-i \frac{\omega}{4\pi T}
\log\left(\frac{1}{\epsilon}\right)
-i\, \Xi 
-i\omega\, \Omega(r_b) 
\right]
\,, 
\end{eqnarray}
where $\Xi$ and $\Omega(r_b)$ are functions resulting from the integral in Eq. \eqref{h_B_UV},  which depends on geometry. The function $\Xi$ is associated with intermediate values of $r$ which is subdominant with respect to the log term and therefore will be ignored in the following. On the other side, $\Omega(r_b)$ depends on the asymptotic behavior of the metric functions for large $r$. 
Close to the boundary, the metric coefficients behave as 
$a(r)\sim r^{a_{t}}\, ,\, b(r)\sim r^{a_{r}}\, ,
\,c(r)\sim r^{a_{x}}\,$, so that the constraint, Eq. \eqref{condUV}, implies that  $\Omega(r_b)$  vanishes in the limit $r_{b} \to \infty$. In particular, if  the metric under consideration is asymptotically AdS, then  $a(r)$, $c(r)$ and $1/b(r)$ go to $r^2$ as $r\to r_b$,  so that $\Omega(r_b) \sim 1/r_b^3$, which is very small for large $r_b$.}


\section{Admittance and diffusion coefficient}\label{sec:Admittance}

In this section we proceed to calculate the admittance from which we find the diffusion coefficient. The admittance is the linear response of the system to an  external force. Thus, first we need to introduce an small external force acting on the boundary particle. With this purpose, we turn on an electromagnetic potential $A_{\mu}$ on the UV brane. This will not change the bulk dynamics but will introduce an external force to the string endpoint. The linearized  Nambu-Goto action, Eq. 
\eqref{ngapprox}, 
then becomes (with unit charge) 
\begin{eqnarray}
\label{eq:actionBH+E}
S 
=- \frac{1}{4 \pi \alpha'} 
\int dt dr 
\left[ \;\dot{X}^2 \frac{g_{xx}
\sqrt{g_{rr}}}{\sqrt{g_{tt}}}
-X'^2\frac{g_{xx}
\sqrt{g_{tt}}}{\sqrt{g_{rr}}} \right]
+ \int dt \left(A_t 
+ \vec{A} \cdot \vec{\dot{x}} \right)\Big|_{r=r_b}\,.\quad 
\end{eqnarray}
From this expression we get a modified  equation of motion \textit{in the brane position} given by 
\begin{eqnarray}
\frac{\sqrt{g_{tt}(r_{b})}}{\sqrt{g_{rr}(r_{b})}}g_{xx}(r_{b})X'|_{r=r_{b}}
-2\pi\alpha'F
=0\,, 
\end{eqnarray}
with $F=\partial_{x} A_{t}-\partial_{t}A_{x}$. 
Since $X'=\partial_{r} X$ and $X(t,r)=e^{i\omega t}h^C_{\omega}(r)$, 
we can write from Eq. \eqref{h_B_UV} 
\begin{eqnarray}
F(\omega)
=\frac{-i\omega  {A_{1}(\omega)}
\sqrt{g_{xx}(r_{h})}}{
2\pi
\alpha'}
f(r_{b})
\approx
\frac{-i\omega  {A_{1}(\omega)}\sqrt{g_{xx}(r_{h})}}{
2\pi
\alpha'}
.
\end{eqnarray}
Note that for this regime of small frequencies and large string energy (large $r_{b}$, $f(r_b) \approx 1$) the force on the particle depends only on the IR structure of the metric. 
Then, the admittance can be written as 
\begin{eqnarray}
\chi(\omega)=\frac{h^C_{\omega}(r_{b})}{F(\omega)}
=2\pi\alpha'
\frac{
\left(1
-i\omega g_{xx}(r_{h})
\int^{r_b}
\frac{
\sqrt{b(r')}}{
g_{xx}(r')f(r')
\sqrt{a(r')}}dr'\right)}{-i
\omega g_{xx}(r_{h})}\,.
\label{general_adm}
\end{eqnarray}
So, its imaginary part is
\begin{eqnarray}
\Im{\chi(\omega)}
=
\frac{2\pi\alpha'}{\omega \, g_{xx}(r_{h})}\,.
\label{general_adm_im}
\end{eqnarray}
The results expressed in Eqs. \eqref{general_adm}-\eqref{general_adm_im} are valid for any metric in the form of Eq. \eqref{gen_metric} in the regime of small frequencies, as proposed in Ref. \cite{Giataganas:2018ekx}. 

From the response function $\chi (\omega)$, Eq. \eqref{general_adm}, we can calculate the diffusion coefficient $D$. It is given by \cite{Tong:2012nf} :
\begin{eqnarray}
\label{Diffusion_Coef}
    D
    =
    \frac{1}{\beta}\lim_{\omega \to 0}
    (-i\omega\chi(\omega))
    =
    \frac{2\pi\alpha'}{\beta g_{xx}(r_{h})}
    =
    \frac{2\pi\alpha' T}{ g_{xx}(r_{h})},
\end{eqnarray}
where $\beta =1/T$. We can rewrite this result just in terms of the metric using \eqref{general_Hawking} as

\begin{equation}
    D=\frac{\alpha' |f'(r_{h})|}{2 g_{xx}(r_{h})}\sqrt{\frac{a(r_{h})}{b(r_{h})}}\,.
\end{equation}

At this stage we have found the linear response and the diffusion coefficient as functions of the black hole radius $r_{h}$. Also we have the temperature as function of the same parameter, so in  principle we can invert the equations to get results as functions of temperature. On the other hand, as temperature depends of different parts of the metric compared with for example admittance we can not write at this point  a general expression for it as function of temperature. We will need then to specialize to some particular model to get such relation.

Before we conclude this section, let us comment that the real part of the admittance $\chi (\omega)$, Eq.  \eqref{general_adm}, is given by
  {
\begin{eqnarray}
\label{real_adm}
\Re\chi(\omega)
&=& 2\pi\alpha'
\int^{r_b}_{r_h+\epsilon}
\frac{
\sqrt{b(r')}}{
g_{xx}(r')f(r') 
\sqrt{a(r')}}dr'\cr 
&\approx & 
\frac{\alpha'}{2Tg_{xx}(r_h)}
\log\left(\frac{1}{\epsilon}\right),
\label{real_adm}
\end{eqnarray}
where we considered the approximation that the integral is dominated by the near horizon (IR) region, as discussed after Eq. \eqref{log}. Note that this expression is independent of the frequencies, at least in the hydrodynamical limit $\omega\to 0$. This result is in consonance with Kramers-Kronig relations such that the real part of the admittance is an even function in~$\omega$.}

\section{Correlation functions}\label{sec:corrfunct}

The mean square displacement is a measure of the variance of the random walk  of the particle in the thermal bath from the motion of the probe string. In the following, we are going to calculate in general grounds this quantity using the results from previous sections. 

First, in order to obtain the mean square displacement one needs to impose ingoing and outgoing boundary conditions near the horizon 
\begin{eqnarray}
\label{General_h_IR}
  h^{IR}_{\omega}(r)&=&
  \frac{  {A(\omega)}}{\sqrt{g_{xx}(r_{h})}}
  \left[e^{i\omega r_{*}}
  +B(\omega)
  e^{-i\omega r_{*}}\right]
  \cr
  &=&\frac{  {A(\omega)}}{\sqrt{g_{xx}(r_{h})}}
  \left[e^{i
  \frac{\omega}{4\pi T} 
\log\left(\frac{r}{r_{h}}
  -1\right)}
  +B(\omega)
  e^{-i\frac{\omega}{4 \pi T} 
  \log\left(\frac{r}{r_{h}}
  -1\right)}\right]\,. 
\end{eqnarray}

In the UV region the solution for small $\omega$ considering these modes is 
\begin{eqnarray}
\label{General_h_UV}
h^{UV}_{\omega}(r)
=\frac{  {A(\omega)}}{
\sqrt{g_{xx}(r_{h})}}
\left[\left(1
-i\omega g_{xx}(r_{h})
\int^{r}_{r_h+\epsilon}
\frac{
\sqrt{b(r')}}{
g_{xx}(r')f(r') 
\sqrt{a(r')}}dr'\right)\right.
\cr
\left.+B(\omega)
\left(1
+i\omega g_{xx}(r_{h})
\int^{r}_{r_h+\epsilon}
\frac{
\sqrt{b(r')}}{
g_{xx}(r')f(r') 
\sqrt{a(r')}}dr'\right)\right] \,. 
\end{eqnarray}
where the coefficients $  {A(\omega)}$ and $  {B(\omega)}$ are the same as in the IR region. 

Matching the UV solution with the Neumann boundary condition in $r=r_{b}$, one obtains that coefficient the $  {B(\omega)}$ is a pure phase   {$e^{i\omega\theta}$ (see the Appendix \ref{sec:neumann})}.  On the other side imposing Neumann b.c. at the IR one obtains a discretization of the frequencies as
\begin{eqnarray}
\label{discreto}
  \Delta\omega
  =\frac{\pi f'(r_{h})}{\log\left(\frac{1}{\epsilon}\right)}
  \sqrt{\frac{a(r_{h})}{b(r_{h})
  }}
  =\frac{4\pi^{2}T}{\log\left(\frac{1}{\epsilon}\right)}\,,
\end{eqnarray}
analogous to the result found in \cite{deBoer:2008gu}. For details, see the Appendix \ref{sec:neumann}. 


\subsection{Grand canonical ensemble and correlation functions}

In the grand canonical ensemble, the density operator is defined as 
\begin{equation}
\rho_{0}=\frac{e^{-\beta(\sum_{\omega>0}\omega a_{\omega}^{\dagger}a_{\omega}-\mu a_{\omega}^{\dagger}a_{\omega}})}{\text{Tr}\left( \, e^{-\beta( H-\mu N)}\right)}\,,    
\end{equation}
where $a_\omega^\dagger$ and $a_\omega$ are the usual creation and annihilation operators which satisfy 
\begin{eqnarray}
\label{ValorEsperado}
\langle a^{\dagger}_{\omega}a_{\omega} \rangle
= \frac{\delta_{\omega\omega'}}{e^{\beta\left(\omega-\mu\right)}\pm 1}; \qquad 
\langle a^{\dagger}_{\omega_n}a^{\dagger}_{\omega} \rangle
=0; \qquad \langle a_{\omega}a_{\omega} \rangle
=0\,. 
\end{eqnarray}
Note that the plus (minus) sign corresponds to the fermionic (bosonic) case. 
With these operators, one can write down the solution for the equation of motion near the boundary 
\begin{eqnarray}
 X(t,r)=\sum_{\omega>0}
 \left(a_{\omega} h_{\omega}^{UV}(r)e^{-i\omega t}+a_{\omega}^{\dagger}(h_{\omega}^{UV}(r))^{*}e^{i\omega t}\right)\,, 
\end{eqnarray}
where 
we used explicitly the quantization of the frequencies obtained in the previous section. Then, the two point function for the string endpoint reads 
\begin{eqnarray}
\label{eq:CorrelationPosition}
\langle x(t)x(0) \rangle &\equiv&
\langle X(t,r_{b})X(0,r_{b}) \rangle
\cr
&=&\sum_{\omega>0}
\sum_{\omega'>0}
\Big( \frac{h^{UV*}_{\omega}(r_{b})
h^{UV}_{\omega'}(r_{b}) e^{i\omega t}
+ h^{UV}_{\omega}(r_{b})
h^{UV*}_{\omega'}(r_{b})
e^{-i\omega  t}}{e^{\beta\left(\omega-\mu\right)} \pm 1}
\cr 
&& \qquad \qquad + h^{UV}_{\omega}(r_{b})
h^{UV*}_{\omega'}(r_{b})
e^{-i\omega  t}\Big)\delta_{\omega\omega'}
\cr
&= & \sum_{\omega>0}
|h^{UV}_{\omega}(r_{b})|^{2}
\Big( \frac{2\cos(\omega t)}{e^{\beta\left(\omega-\mu\right)} \pm 1}
+ e^{-i\omega  t}\Big). 
\end{eqnarray}

Substituting in the above equation the expression for $h_\omega^{UV}(r_b)$ given by \eqref{General_h_UV},   { taking into account the fact that $A(\omega)\sim \omega^{-1/2}$ from Eq. \eqref{NormA}}, and disregarding terms of order  $\omega^{2}$, we obtain
\begin{eqnarray}
\langle x(t)x(0) \rangle
&=&\frac{
4\pi^2  \alpha' T}{
g_{xx}(r_{h})
\log\left(\frac{1}{\epsilon}\right)}
\sum_{\omega>0}
\frac{1}{\omega}
\left( \frac{2\cos(\omega t)}{
e^{\beta\left(\omega-\mu\right)}\pm 1}
+ e^{-i\omega  t}\right)\,.
\end{eqnarray}
Considering the approximation
\begin{eqnarray*}
    d\omega\sim
    \Delta\omega
    =\frac{4 \pi^2 T}{
\log\left(\frac{1}{\epsilon}\right)},
\end{eqnarray*}
one can rewrite
\begin{eqnarray}
\label{2Point}
\langle x(t)x(0)\rangle
&=&\frac{
\alpha'}{
g_{xx}(r_{h})}
\int_{0}^{\infty}
\frac{d\omega}{\omega}
\left( \frac{2\cos(\omega t)}{
e^{\beta\left(\omega-\mu\right)}\pm 1}
+ e^{-i\omega  t}\right)\cr \cr 
&=&  \langle x(0)x(t)\rangle^*\, .
\end{eqnarray}
Analogously, one can obtain 
\begin{eqnarray}
\label{eq:CorrelationPositionTime}
\langle x(t)x(t)\rangle
&=& \langle X(t,r_{b})X(t,r_{b}) \rangle
\nonumber \\
&=& \sum_{\omega>0}
|h^{UV}_{\omega}(r_{b})|^{2}
\Big( \frac{2}{e^{\beta(\omega-\mu)}\pm 1}
+ 1\Big)
\nonumber \\
&=& \frac{\alpha' }{g_{xx}(r_{h})}
\int_{0}^{\infty}
\frac{d\omega}{\omega}
\left( \frac{2}{e^{\beta(\omega-\mu)}\pm 1}
+ 1\right)
\cr \cr 
&=& \langle x(0)x(0) \rangle\,.
\end{eqnarray}

  {Note that in Eqs. \eqref{eq:CorrelationPosition}- \eqref{eq:CorrelationPositionTime} the sums and integrals span over all positive frequencies $\omega$. On the other side, 
the modes present in 
$|h^{UV}_{\omega}(r_{b})|^{2}$ are the first terms of an expansion for $\omega/T \ll 1$, as discussed after Eq.\eqref{h_B_UV}. As we show in Appendix \ref{energy_scale}, 
when we substitute  $|h^{UV}_{\omega}(r_{b})|^{2}$ by Eq. \eqref{General_h_UV} in the above equations, the frequencies 
$\omega \gg T $ are exponentially suppressed. Then,  it is a good approximation to keep only the first terms in an expansion of $\omega/T $ for $|h^{UV}_{\omega}(r_{b})|^2$.}


\section{Fluctuation-dissipation theorem} \label{sec:flutuation-dissipation}

In this Section, we are going to verify the consistency of our results with the fluctuation-dissipation theorem, using the admittance obtained in Section  \ref{sec:Admittance} and the correlation functions in Section \ref{sec:corrfunct}.

Starting from the correlation functions $\langle x(t)x(0) \rangle$ and $\langle x(0)x(t) \rangle$,  one can define a symmetric Green's function  as:
\begin{equation}\label{Gsim}
 G_{\rm Sym}(t)\equiv \frac{1}{2}\left(\langle x(t)x(0) \rangle+\langle x(0)x(t) \rangle\right)\,. 
\end{equation}
The fluctuation-dissipation theorem  in the presence of a chemical potential for the bosonic \cite{Zubarev} and fermionic \cite{Markov:2009ue} cases, can be written as 
\begin{eqnarray}
\label{FDT} 
    G_{\rm Sym}^{B,F}(t) =\mathcal{F}^{-1}\left[\left(1+2n_{B,F}\right)\Im \chi (\omega)\right],
\end{eqnarray}
where $\mathcal{F}^{-1}[\cdots]$ denotes the inverse Fourier transform, $n_{B,F}$ are the Bose-Einstein and Fermi-Dirac distributions,  and $\Im \chi(\omega)$ is the imaginary part of the admittance. Note that the frequencies in this equation are positive physical quantities. 
 
 Then, one can rewrite of the  r.h.s of equation \eqref{FDT} as Fourier transforms as
\begin{eqnarray}
\label{Vascao}
\mathcal{F}^{-1}\left[\left(1+2n_{B,F}\right)\Im \chi (\omega)\right]  &=& \frac{1}{2\pi}\int_{-\infty}^{\infty}d\omega\left(1+\frac{2}{e^{\beta(|\omega|-\mu)}\pm 1}\right)\Im \chi (\omega)e^{i\omega t} \cr 
    &=& \frac{\alpha' }{g_{xx}(r_{h})}\int_{-\infty}^{\infty}\frac{d\omega}{|\omega|}\left(1+\frac{2}{e^{\beta(|\omega|-\mu)}\pm 1}\right)e^{i\omega t}\,, 
\end{eqnarray}
where we used the admittance, Eq. \eqref{general_adm_im},  and 
the plus (minis) sign represents fermions (bosons).

On the other hand, from the correlation functions, Eqs. \eqref{2Point},  the l.h.s. of Eq. \eqref{Gsim} becomes 
\begin{eqnarray}
G_{\rm Sym}^{B,F}(t)&=&
\frac{\alpha' }{g_{xx}(r_{h})}\int_{0}^{\infty}\frac{d\omega}{\omega}
\left( \frac{4\cos(\omega t)}{e^{\beta\left(\omega-\mu\right)}\pm 1}
+ e^{-i\omega  t}+e^{i\omega  t}\right)\cr
&=&\frac{\alpha' }{g_{xx}(r_{h})}\int_{0}^{\infty}\frac{d\omega}{|\omega|}
\left( \frac{2\left( e^{-i\omega  t}+e^{i\omega  t}\right)}{e^{\beta\left(|\omega|-\mu\right)}\pm 1}
+ e^{-i\omega t}+e^{i\omega  t}\right).
\label{SymG}
\end{eqnarray}
Since 
\begin{eqnarray}
   \int_{0}^{\infty}
   d\omega f(|\omega|)
   e^{i\omega t}
   +\int_{0}^{\infty}d\omega f(|\omega|)
   e^{-i\omega t}
   =\int_{-\infty}^{\infty}d\omega f(|\omega|)
   e^{i\omega t}\,, 
\end{eqnarray}
the Eq.\eqref{SymG} becomes
\begin{eqnarray}
G_{\rm Sym}
&=&\frac{\alpha' }{g_{xx}(r_{h})}
\int_{-\infty}^{\infty}\frac{d\omega}{|\omega|}
\left( \frac{2 }{e^{\beta\left(|\omega|-\mu\right)}\pm 1}
+1\right) e^{i\omega  t}\,, 
\end{eqnarray}
which coincides with Eq. \eqref{Vascao}. Therefore, it completes the check of our calculations with the fluctuation-dissipation theorem.


\section{Mean square displacements} \label{sec:msd}

From the results of Section  \ref{sec:corrfunct}, one can calculate the regularized expression for the mean square displacement as: 
\begin{equation}
    s^2_{\rm reg}(t)
    = \langle
    : [x(t)-x(0)]^2 : \rangle
    \equiv \langle : 
    [X(t, r_b)-X(0, r_b)]^2 : \rangle\,,
\end{equation}
where the double dots notation $:  [\cdots] :$ means that we are using  normal ordering. 
Explicitly, 
\begin{eqnarray}
\label{sreggeral}
s^2_{\rm reg}(t)
&=&\frac{\alpha' }{g_{xx}(r_{h})}
\int_{0}^{\infty}
\frac{d\omega}{\omega}
\frac{4(1-\cos(\omega t))}{e^{\beta(\omega-\mu)}\pm 1}
\cr
&=&\frac{8\alpha' }{g_{xx}(r_{h})}
\int_{0}^{\infty}
\frac{d\omega}{\omega}
\frac{\sin^{2}\left(\frac{\omega t}{2}\right)}{e^{\beta(\omega-\mu)}\pm 1}.
\end{eqnarray}
In the following, we are going to calculate the regularized mean square displacement in various interesting cases. 

\subsection{Zero chemical potential }

In this section, we study the particular case of
zero chemical potential from previous results. This is interesting, for instance, in models which describe the superfluid Bose-Mott insulator transition \cite{Zaanen:2015oix} and massless Dirac fermions in graphene \cite{novoselov,moriconi}.

For zero chemical potential, the mean square displacement, Eq. \eqref{sreggeral}, reads
\begin{eqnarray}
\label{zerochemical}
s^2_{\rm reg}(t)
&=&\frac{8\alpha' }{g_{xx}(r_{h})}
\int_{0}^{\infty}
\frac{d\omega}{\omega}
\frac{\sin^{2}\left(\frac{\omega t}{2}\right)}{e^{\beta\omega}\pm 1}\,. 
\end{eqnarray}
This expression is valid for the bosonic $(-)$ and fermionic $(+)$ cases. In the following we discuss separately these two cases. 

\subsubsection{Bosons}
 
To solve the integral \eqref{zerochemical} for the bosonic case $(-)$, we follow  \cite{Caldeira:2020sot} and find 
\begin{eqnarray}
\label{bosonic_zeromu}
s^2_{\rm reg}(t)
&=&
\frac{2\alpha' }{g_{xx}(r_{h})}
\log \left(\frac{\sinh (\frac{t \pi}{\beta})}{\frac{t \pi}{\beta}} \right) \,. 
\end{eqnarray}

For the short time approximation $t\ll \beta$, we have 
\begin{eqnarray}
s^2_{\rm reg}(t) 
\approx 
\frac{
\pi^{2}\alpha'}{
3g_{xx}(r_{h})}
\frac{
t^2 
}{
\beta^2}\,,
\end{eqnarray}
which is the expected result for the ballistic regime. 
Considering now the approximation for large times $t\gg \beta$, one finds 
\begin{eqnarray}
s^2_{\rm reg}(t) 
\approx 
\frac{
2\pi\alpha'}{
g_{xx}(r_{h})}
\frac{
t 
}{
\beta}\,, 
\end{eqnarray}
which is the standard diffusion result.

\subsubsection{Fermions}
 
\label{Fermi_zero_mu} 
Now, considering the fermionic case $(+)$ corresponding to the integral \eqref{zerochemical}, we have 
\begin{eqnarray*}
s^2_{\rm reg}(t)
&=&\frac{8\alpha' }{
g_{xx}(r_{h})
}
\int_{0}^{\infty}
\frac{d\omega}{\omega}
\frac{
\sin^{2}\left(\frac{\omega t}{2}\right)}{
e^{\beta\omega}+1
}\,. 
\label{int_fermion}
\end{eqnarray*}
  {
\begin{eqnarray}
&=&
\frac{2\alpha' }{
g_{xx}(r_{h})
}
\log \left(\frac{\frac{t\pi}{2\beta}}{\tanh\left(\frac{\pi t}{2\beta}\right)}\right)\, .
\end{eqnarray}
}

  {In the limit $t \ll \beta$ (short times) one can approximate this expression by
\begin{eqnarray}
\label{fermion_short_zeromu}
s^2_{\rm reg}(t)
&\approx&
\frac{\pi^2 \alpha'}{6 g_{xx}(r_{h})}
\frac{t^{2}}
{\beta^{2}}\, ,   
\end{eqnarray}
in agreement with the well known result for the ballistic regime.}

  {
For large times, or $t \gg \beta$, we have that
\begin{eqnarray}
\label{fermion_long_zeromu}
s^2_{\rm reg}(t)
&\sim&
\frac{2\alpha' }{
g_{xx}(r_{h})
}
\log \left(\frac{\pi t}{2\beta }\right)\,,
\end{eqnarray}
which is a Sinai-like subdiffusive regime \cite{sinai-like}}.

\subsection{Finite chemical potential }

In the case of a finite chemical potential the mean square displacement is given by Eq. \eqref{sreggeral} which we repeat here for convenience 
\begin{eqnarray}
\label{sreggeral2}
s^2_{\rm reg}(t)
&=&\frac{8\alpha' }{g_{xx}(r_{h})}
\int_{0}^{\infty}
\frac{d\omega}{\omega}
\frac{\sin^{2}\left(\frac{\omega t}{2}\right)}{e^{\beta(\omega-\mu)}\pm 1}\,.
\end{eqnarray}
In the following, we specialize to the  bosonic $(-)$ and fermionic $ (+)$ cases. 

\subsubsection{Bosons}

In order to calculate the mean square displacement given by the above equation  
for the bosonic case, we take $\mu < 0$, and consider the series expansion 
\begin{equation}\label{eq:geometricseries}
   \frac{1}{e^{\beta( \omega-\mu)}- 1}  =  \frac{e^{-\beta( \omega-\mu)}}{1- e^{-\beta( \omega-\mu)}} = \sum_{n = 0}^{\infty}e^{-\beta (\omega-\mu) (n+1)}\,.
\end{equation}
Then, 
\begin{equation}
s^2_{\rm Breg}(t) =\frac{8\alpha' }{g_{xx}(r_{h})}\sum_{n=1}^{\infty}\int_{0}^{\infty}\frac{d\omega}{\omega}e^{-\beta (\omega-\mu) n}\sin^{2}(\frac{\omega t}{2}).
\end{equation}
Performing the integral, one gets 
\begin{equation}
\label{S2Series}
s^2_{\rm Breg}(t) =\frac{2\alpha' }{g_{xx}(r_{h})}\sum_{n=1}^{\infty}e^{\beta\mu n}\log \left(1 + \frac{t^2}{n^2 \beta^2} \right)\,.
\end{equation}
This expression can be rewritten in a formal way as   
\begin{eqnarray}
    s^{2}_{\rm Breg}(t)
    =
    \frac{2\alpha' }{g_{xx}(r_{h})}
    \left\{2 {\rm Li_{0}}^{(1,0)}\left(0,e^{\mu }\right) - e^{\mu } \left[{\Phi}^{(0,1,0)}\left(e^{\mu },0,1+i \frac{t}{\beta}\right)
    \right.\right. 
    \cr
    \left.\left. 
    +{\Phi}^{(0,1,0)}\left(e^{\mu },0,1-i \frac{t}{\beta}\right)\right]\right\}\,. 
\end{eqnarray}
In this equation we used the following notation: Li$_n^{(1,0)}(x,y)$ is the first derivative of the polylogarithm function of order $n$ with respect to its first argument $x$;  $\Phi^{(0,1,0)}(x,y,z)$ is the first derivative of the Lerch transcendent function with respect to the second argument $y$.

First we consider $t\ll\beta$, which is the short time approximation, and then from Eq. \eqref{S2Series}, we find
\begin{eqnarray}\label{bostpeq}
     s^{2}_{\rm Breg}(t)
     &\approx&
     \frac{2\alpha' }{g_{xx}(r_{h})}
     \left(\sum _{n=1}^{\infty } \frac{e^{\beta\mu  n}}{n^2}\right)\frac{t^{2}}{\beta^{2}}
     =
     \frac{2\alpha' }{g_{xx}(r_{h})}
     \text{Li}_2\left(e^{\beta\mu }\right)\frac{t^{2}}{\beta^{2}}
     \,. 
\end{eqnarray}
In the right hand side of this equation we used the polylogarithm function of order 2, $\text{Li}_2\left(e^{\beta\mu }\right)$. This equation gives the usual ballistic behavior since it goes like $ t^2$.

  {
On the other side, in the late time approximation $t\gg\beta$ the sum in Eq. \eqref{S2Series} is dominated by its first term and we have
\begin{eqnarray}
\label{bostgran}
s^{2}_{\rm Breg}(t)
&\approx&\frac{4\alpha' }{g_{xx}(r_{h})}
e^{\beta\mu}\log \left( \frac{t}{ \beta} \right)\,. 
\end{eqnarray}
}
This equation corresponds to a subdiffusive regime which is due to the presence of the non-zero chemical potential in this case. This result is analogous to what was found in classical physical systems in \cite{sinai-like} or in Lorentz invariant bosonic theories 
\cite{Caldeira:2021izy}.

\subsubsection{Fermions}

 Now, we consider the regularized mean square displacement for the fermionic case. From Eq. \eqref{sreggeral},  with $\mu > 0$, we have 
\begin{equation}
        \label{sfreg}
     s^2_{\rm Freg}(t)=\frac{8\alpha' }{g_{xx}(r_{h})}\int_{0}^{\infty}\frac{d\omega}{\omega}\left( \frac{\sin^{2}(\frac{\omega t}{2})}{e^{\beta(\omega-\mu)}+1}\right)\,. 
\end{equation} 
In order to evaluate this integral we resort to the Sommerfeld expansion:  
\begin{eqnarray}
\int_{0}^{\infty}\frac{d\omega}{\omega}\left( \frac{\sin^{2}(\frac{\omega t}{2})}{e^{\beta(\omega-\mu)}+1}\right)&=&\int_{0}^{\mu}\frac{\sin^{2}(\frac{\omega t}{2})}{\omega}d\omega+\cdots 
\cr
&=&\frac{1}{2} (-\text{Ci}(t \mu )+\log (t\mu )+\gamma )+\cdots, 
\end{eqnarray}
where $\text{Ci} (z)$ is the Cosine integral function, $\gamma$  is Euler-Mascheroni constant and we disregarded terms of order ${1}/{\mu^{2}\beta^{2}}$ and higher, since we are considering the low temperature regime,  $\mu\beta=\frac{\mu}{T}\gg1$.

The mean square displacement, Eq. \eqref{sfreg}, for small times $\mu t \ll 1$, can be approximated as
\begin{equation}\label{fertpeq}
    s^2_{\rm Freg}(t)
    \approx
    \frac{\alpha' }{g_{xx}(r_{h})}\mu^{2}t^2\,. 
\end{equation}
This expression for $s^2$ gives  the well known ballistic regime $ t^2$.

On the other hand, taking the large time approximation $ \mu t \gg 1$ in Eq. \eqref{sfreg}, we obtain 
\begin{equation}\label{fertgran}
     s^2_{\rm Freg}(t) \approx
     \frac{4\alpha' }{g_{xx}(r_{h})} \log \left(t\mu \right),
\end{equation}
which corresponds to a subdiffusive behavior $ \log t$, as in the bosonic case discussed above. Note that this behavior also appears in classical  set-ups as in Ref.  \cite{sinai-like}. In Ref. \cite{Caldeira:2021izy} we found a similar fermionic behavior for a Lorentz invariant context. 
 
\subsection{Summary and discussions on $s_{\rm reg}^2$} 
 
We can now summarize the  results of this section on Table \ref{table1}.    
First, one can note that for all scenarios we get the usual ballistic regime $s^{2}_{\rm reg}\sim t^{2}$ for short times. This is expected since for this situation the particle do not had completely felt the characteristics of the environment. Another feature for all these results is the proportionality to the inverse of $g_{xx}(r_{h})$, showing the dependence on the IR of the metric in the direction of motion of the test particle.

For the bosonic case, the short time behavior at zero chemical potential is obtained directly just taking the $\mu=0$ limit from the finite chemical potential result. However, at large times, one can not obtain the zero chemical potential case by taking the above limit. In this case it is necessary to go back to Eq. \eqref{sreggeral} and recalculate this quantity.  Then turning on a chemical potential changes the long time behavior of the regularized mean square displacement for the bosonic scenario.   

In doing the present analysis it is necessary to bear in mind that the term $g_{xx}(r_{h})$ will bring a non-trivial dependence on the temperature and on the  chemical potential. So, in general, the temperature behavior of these quantities depends heavily on the particular form of the metric. In the following sections we apply the results obtained here for two set ups as a illustration for those aspects.

As a general comment, it is interesting to note that the set up discussed above for bosons and fermions at zero or non-zero chemical potential verifies the fluctuation-dissipation theorem. This is done relating the imaginary part of the admittance with the two point correlation functions calculated above, as discussed in Section 
 \ref{sec:flutuation-dissipation}.

\begin{table}
\begin{center}
\begin{tabular}{||c | c | c ||}
\hline
Description & $s^{2}_{\text{reg}}(t)$ short times & $s^{2}_{\text{reg}}(t)$ large times
 \T\B \\ 
\hline \hline
Bosonic $\mu=0$ &
$\frac{
\pi^{2}\alpha'}{
3g_{xx}(r_{h})}
\frac{
t^2 
}{
\beta^2}
$
& 
$
\frac{
2\pi\alpha'}{
g_{xx}(r_{h})}
\frac{
t 
}{
\beta}
=
D t $
\T\B \T\B
\\
\hline
Bosonic $\mu\not=0$
&
$\frac{2\alpha' }{g_{xx}(r_{h})}
\text{Li}_2\left(e^{\beta\mu }\right)\frac{t^{2}}{\beta^{2}}$
& 
  {
$\frac{4\alpha' }{g_{xx}(r_{h})}
e^{\beta\mu}\log \left( \frac{t}{ \beta} \right)
=
\frac{2\beta D}{\pi}e^{\beta\mu}\log \left( \frac{t}{ \beta} \right)$} \T\B \T\B
\\ 
\hline
Fermionic $\mu=0$ 
& 
$\frac{\pi^2 \alpha'}{6 g_{xx}(r_{h})}
\frac{t^{2}}
{\beta^{2}}
$
&
  {
$\frac{2\alpha' }{
g_{xx}(r_{h})
}
\log \left(\frac{\pi t}{2\beta }\right)
=
\frac{\beta D}{\pi}
\log \left(\frac{\pi t}{2\beta }\right)$ 
}\T\B \T\B
\\
\hline
Fermionic $\mu\not=0$
&
$\frac{\alpha' }{g_{xx}(r_{h})}\mu^{2}t^2
$
& $\frac{4\alpha' }{g_{xx}(r_{h})} \log \left(t\mu \right)
=
\frac{2 \beta D}{\pi} \log \left(t\mu \right)$ \T\B \T\B
\\
\hline
\end{tabular}
\end{center}
\caption{Mean square displacement for zero and non-zero chemical potential for bosons and fermions.}
\label{table1}
\end{table} 
%


\section{Hyperscaling  violation at finite temperature for bosons and fermions}\label{sec:hyp}

In this section, we apply the general results obtained in the previous parts of the text to a particular system, the Lifshitz-hyperscaling family of metrics:
\begin{eqnarray}
\label{Lifshitz_Metric}
ds^{2}
=
r^{-\frac{2\theta}{d}}
\left(
-r^{2z}f(r)dt^{2}+r^{2}d\vec x^{2}+\frac{dr^{2}}{f(r)r^{2}}
\right),
\end{eqnarray}
where $z$ and $\theta$ are the Lifshitz (or dynamical) and hyperscaling violation  parameters respectively, and $d$ is the spacetime dimension. The horizon function is given by
\begin{eqnarray}
\label{lf}
f(r)=1-\left(\frac{r_{h}}{r}\right)^{d+z-\theta}.
\end{eqnarray}

This problem was studied by the authors in Refs. \cite{Edalati:2012tc}  and \cite{Giataganas:2018ekx} for some particular cases without fermions. In our presentation we will discuss the Lifschitz-hyperscaling violation metrics at finite temperature for fermions and bosons at zero chemical potential.

  {The constraint in equation \eqref{condUV} applied for the hyperscaling-Lifshitz metric case becomes 
\begin{eqnarray}
    \frac{2 \theta }{d}+ z+1 \geq 0
\end{eqnarray}
It is also important to note as pointed in \cite{Dong:2012se}  that the null energy condition imposes additional constraints in the values of the parameters $\theta$ and $z$, explicitly
\begin{eqnarray}
    (d-\theta)
    (d(z-1)-\theta)
    \geq 0,
    \cr
    (z-1)(d+z-\theta)
    \geq 0.
\end{eqnarray}
In order to have a positive specific heat we need to impose one more condition  \cite{Dong:2012se}
\begin{eqnarray}
\label{Calor_esp}
\frac{d-\theta}{z}\geq 0\,.
\end{eqnarray}
Considering that the number $d$ of spatial dimensions in the boundary  is a positive number and imposing all these conditions results in the parameters assuming the possible values
}
  {
\begin{eqnarray}
&(i)& \quad z\leq -3,\,\,
\theta \geq -\frac{d}{2} (z+1), \\
&(ii)& \quad -3<z<0,\,\, \theta \geq d,
\\
&(iii)& \quad 1\leq z<2,\,\,
-\frac{d}{2} ( z+1)\leq \theta \leq d( z-1), 
\,\,\text{or} \,\, \theta =d, 
\\
&(iv)&\quad z\geq 2,\,\, -\frac{d}{2} (z+1)\leq \theta \leq d.
\end{eqnarray}
}
In this background, we then calculate the response function, the diffusion coefficient, and the mean square displacement from our general discussion of the previous sections. This will allow us to understand the independence of the diffusion process on different hyperscaling parameters, dynamical exponents, and dimensions for bosons and fermions. Also, it will be illustrative to explore some particular configurations of those quantities, presenting some cases in detail. Such Analysis is interesting since those types of gravitational systems are relevant as holographic duals for some  condensed matter systems. In some works, for example, they are seen as possessing Fermi surfaces in the boundary theory such as strange metals  \cite{Huijse:2011ef,Zaanen:2015oix, Hartnoll:2018xxg,Gubser:2000mm,Dong:2012se,Ogawa:2011bz,Kachru:2008yh}. In other cases, they were used to study the behavior of some (bosonic) theories close to critical point \cite{Edalati:2012tc,Tong:2012nf}.

The horizon function Eq. \eqref{lf} implies that the Hawking temperature reads 
\begin{eqnarray}
T
=
\frac{
\left| d+z-\theta \right| }{
4 \pi }\, r_{h}^{z}\;. 
\end{eqnarray}
Then, we can write the admittance from Eq. \eqref{general_adm_im} as 
\begin{eqnarray}
\Im{\chi}(\omega)
= 
\frac{2\pi\alpha'}{\omega }
\left(
\frac{4 \pi}{
\left| d+z-\theta \right| }
\right)^{-{2( d-\theta)}/{zd}}
T^{-{2( d-\theta)}/{zd}}.
\label{admitt_hyp}
\end{eqnarray}
Note that the condition \eqref{Calor_esp} implies a negative exponent for the temperature dependence of the imaginary part of the admittance, which is the usual behavior for holographic systems such as the ones investigated in Ref. \cite{deBoer:2008gu}. 

Now, using Eq. \eqref{Diffusion_Coef}, the diffusion coefficient for this model is  
\begin{eqnarray}
D=2\pi\alpha'
\left(
\frac{4 \pi}{
\left| d+z-\theta \right| }
\right)^{-{2( d-\theta)}/{zd}}
T^{(-{2( d-\theta)+zd)}/{zd}}.
\label{diffus_hyp}
\end{eqnarray}
 It is interesting to note that this diffusion coefficient could increase or decrease with the temperature depending on the choice of the parameters $z, d, \theta$.

In the particular case where $d=\theta$ one has the usual Einstein diffusion proportional to $T$. On the other side, for $ d-\theta=zd$ one has the diffusion coefficient proportional to $T^{-1}$. This behavior was also found in Ref. \cite{deBoer:2008gu} for pure AdS space. These  cases are thermodynamically stable accordingly to the condition given by Eq. \eqref{Calor_esp}.

\subsection{Bosons}

The regularized mean square displacement for the bosonic case without chemical potential ($\mu=0$), Eq. \eqref{bosonic_zeromu},
on a Lifshitz and hyperscaling violation metric reduces to 
\begin{eqnarray}
s^2_{\rm Breg}(t)
&=&
{2\alpha' }{ \left(\frac{4\pi T}{| d+z-\theta | }\right)^{-{2( d-\theta)}/{zd}}}
\log \left(\frac{\sinh ({T t \pi})}{{T t \pi}} \right)\,.
\label{sbreg_hl}
\end{eqnarray}
For short times one reobtains  the ballistic regime
\begin{eqnarray}
s^2_{\rm Breg}(t)
&=&
\frac{2\alpha' }{3}
{ \left(
\frac{4\pi T}{| d+z-\theta | }\right)^{-{2( d-\theta)}/{zd}}}
(T t \pi)^2\,
\label{s_ballis_Boson}
\end{eqnarray}
 and for long times we find
\begin{eqnarray}
s^2_{\rm Breg}(t)
&=& 
{
2\alpha' }{\left(\frac{4 \pi T}{| d+z-\theta | }\right)^{- \frac{2}{d}
\frac{d- \theta }{z}} }
\left(
T t \pi 
\right)\cr 
&=& 
{
2\pi \alpha' }{\left(\frac{4 \pi }{| d+z-\theta | }\right)^{- \frac{2}{d}
\frac{d- \theta }{z}} }
\left(
T 
\right)^{- \frac{2}{d}
\frac{d- \theta }{z}+1}\, t
\end{eqnarray}
which is the usual diffusive  behavior in time.

\subsection{Fermions}

Now, we are going to analyze  the fermionic case with zero chemical potential within a Lifshitz and hypescaling violation metric. So, here we will calculate the mean square displacement for this system. We get from Eq.  \eqref{fermion_short_zeromu} that for short times one has 
\begin{eqnarray}
s^2_{\rm Freg}(t)
\approx
\frac{2\alpha'}{
g_{xx}(r_{h})
}
\frac{t^{2}}{\beta^{2}}
=
2\alpha'
\left(\frac{4\pi T}{| d+z-\theta | }\right)^{-{2( d-\theta)}/{zd}}
\frac{t^{2}}{\beta^{2}}\,,
\label{s_Freg_hyp_s}
\end{eqnarray}
which is the typical ballistic behavior while for long times, in turn, from Eq.  \eqref{fermion_long_zeromu}, one finds 
\begin{eqnarray}
s^2_{\rm Freg}(t)
&\sim&
\frac{2\alpha' }{
g_{xx}(r_{h})
}
\log \left(\frac{\pi t}{2 \beta }\right)
=
2\alpha'
\left(\frac{4\pi T}{| d+z-\theta | }\right)^{-{2( d-\theta)}/{zd}}
\log \left(\frac{\pi t}{ 2 \beta }\right)
\,, \label{s_Freg_hyp_l}
\end{eqnarray}
which corresponds to a fermionic subdiffusive regime analogous to the classical (Boltzmann) ones found in Ref. \cite{sinai-like}. 

Now, we are going to discuss two special cases: $\theta=d$, which reproduces the Einstein diffusion coefficient, and $\theta=d-1$, in which the admittance goes with the inverse temperature as in the pure AdS case,  in the next two sections.

\subsection{The particular case $\theta=d$}

In this section, we discuss the Lifshitz and hyperscaling violation at zero chemical potential ($\mu=0$) presented above for the particular case of $\theta=d$.  
In this case,  the admittance, Eq. \eqref{admitt_hyp}, is independent of the temperature
\begin{eqnarray}
\chi(\omega)|_{\theta=d}
=
\frac{2\pi \alpha'  }{\omega }\,,
\end{eqnarray}
and the diffusion coefficient, Eq. \eqref{diffus_hyp}, becomes 
\begin{eqnarray}
D|_{\theta=d}=2\pi\alpha' T, 
\end{eqnarray}
which is the usual expected behavior for Brownian motion, as obtained by Einstein in his original formulation of the problem. 

\subsubsection{Bosons}

The regularized mean square displacement for bosons, Eq. \eqref{sbreg_hl}, in this case is given by 
\begin{eqnarray}
s^2_{\rm Breg}(t)|_{\theta=d}
&=&
{2\alpha' }
\log \left(\frac{\sinh ({T t \pi})}{{T t \pi}} \right)\,.
\end{eqnarray}
Taking this expression in the limit of short times, one finds  the usual ballistic regime 
\begin{eqnarray}
s^2_{\rm Breg}(t)|_{\theta=d}
&=&
\frac{2\alpha' }{3} 
(T t \pi)^2,
\end{eqnarray}
while for long times one finds 
\begin{eqnarray}
s^2_{\rm Breg}(t)|_{\theta=d}
= 
2\pi\alpha' 
T t
=
Dt\,,
\end{eqnarray}
which is also the usual Einstein diffusive regime. 

\subsubsection{Fermions}

In this case ($\theta =d$) the fermionic mean square displacement from \eqref{s_Freg_hyp_s}  for small times reads the natural ballistic profile  
\begin{eqnarray}
s^2_{\rm Freg}(t)|_{\theta=d}
\approx
2\alpha'
\frac{t^{2}}{\beta^{2}}\,.
\end{eqnarray}
From the diffusive regime \eqref{s_Freg_hyp_l}, we get
\begin{eqnarray}
s^2_{\rm Freg}(t)|_{\theta=d}
\approx
2\alpha'
\log \left(\frac{\pi t}{2\beta }\right), 
\end{eqnarray}
which is also a subdiffusive behavior as the ones found in Ref. \cite{sinai-like}.

It is interesting to note that in this case $\theta =d$ there is no dependence on the dynamical exponent $z$ in the admittance, although the diffusion coefficient and the mean square displacement depend on the temperature.

In ref. \cite{Dong:2012se} the authors  consider $d-\theta$ as an effective spatial dimension so that one obtains a $0+1$ dimensional system regarding the entropy of the system. However, even for $\theta=d$ one finds non zero spatial correlations, as we found above.

\subsection{The particular case $\theta=d-1$}

The particular case $\theta=d-1$ is interesting because it can be related to a compressible state  with hidden Fermi surfaces, as discussed in Ref. \cite{Huijse:2011ef}.
In this case without chemical potential, the imaginary part of the admittance, Eq. \eqref{admitt_hyp}, 
gives 
\begin{eqnarray}
\Im{\chi}(\omega)|_{\theta=d-1}
= 
\frac{2\pi\alpha'  }{\omega }\left( \frac{4\pi T}{\left| z+1 \right| }\right)^{-{2}/{zd}}
\end{eqnarray}
By the condition \eqref{Calor_esp} which keeps the specific heat positive one finds that in this case one has $z>0$, so that the imaginary part of the admittance is proportional to some inverse power of the temperature. It is also remarkable that for $z\to \infty$ the imaginary part of the admittance becomes a constant independent of the temperature.  

The diffusion coefficient, Eq. \eqref{diffus_hyp}, will be now
\begin{eqnarray}
D|_{\theta=d-1}=2\pi\alpha'
\left(
\frac{4 \pi}{
\left| z+1 \right| }
\right)^{-{2}/{zd}}
T^{1-2/zd}.
\label{diffus_hyp2}
\end{eqnarray}
If $z\to \infty$ this diffusion coefficient becomes proportional to $T$, and if $zd=2$ it becomes a constant. 

\subsubsection{Bosons}

From Eq. \eqref{sbreg_hl}
one finds for $\theta=d-1$
\begin{eqnarray}
s^2_{\rm Breg}(t)
&=&
{2\alpha' }{ \left(\frac{4\pi T}{| z+1 | }\right)^{-{2}/{zd}}}
\log \left(\frac{\sinh ({T t \pi})}{{T t \pi}} \right)\,.
\end{eqnarray}
Then, for short times one has the ballistic regime
\begin{eqnarray}
s^2_{\rm Breg}(t)|_{\theta=d-1}
&=&
\frac{2\alpha' }{3}
{ \left(\frac{4 }{| z+1| }\right)^{- {2}/{zd}
} }
\left(
\pi T  
\right)^{ 2-{2}/{zd}}
t^2\,
\end{eqnarray}
and for long times the diffusive response 
\begin{eqnarray}
s^2_{\rm Breg}(t)|_{\theta=d-1}
&=& 
{
2\alpha' }{\left(\frac{4 }{| z+1| }\right)^{- \frac{2}{dz}
} }
\left(
\pi T  
\right)^{- \frac{2}{dz}+1}t.
\end{eqnarray}
Note that for $z>2/d$ the mean square displacement is proportional 
to $T^{\alpha}$ with $\alpha >0$, in both regimes. If $z\to \infty$, then the ballistic regime goes with $T^2$ and the diffusion as $T$. 

\subsubsection{Fermions}

 For fermions in this special case and for short times the mean square displacement, 
Eq.   \eqref{s_Freg_hyp_s}, becomes
\begin{eqnarray}
s^2_{\rm Freg}(t)|_{\theta=d-1}
\approx
2\alpha'
\left(\frac{4\pi T}{| z+1 | }\right)^{-{2}/{zd}}
\frac{t^{2}}{\beta^{2}}\,.
\end{eqnarray}

In the diffusive regime (long   times) we have, from Eq.  \eqref{s_Freg_hyp_l}, that 
\begin{eqnarray}
s^{2}_{\rm Freg}|_{\theta=d-1}
=
2\alpha'
\left(\frac{4\pi T}{| z+1 | }\right)^{-{2}/{zd}}
\log \left(\frac{\pi t}{2 \beta }\right)
\,.
\end{eqnarray}

\section{Charged dilatonic AdS black holes}
\label{sec:GRM} 

Our objective here is provide an application of ours findings to a non zero chemical potential system. As an example, we consider the Gubser-Rocha model \cite{Gubser:2009qt,Zaanen:2015oix}, which is a top-down construction from eleven to four spacetime dimensions for a charged dilatonic AdS black hole. In this model there are fermionic modes which indicate the presence of Fermi surfaces.  The lagrangian in four dimensions is given by
\begin{eqnarray}
{\cal L}
= 
R -\frac{1}{4} e^{\phi/\sqrt{3}}F_{\mu\nu}^2 
-\frac{1}{2}(\partial_\mu\phi)^2 
-\frac{6}{L^2}\cosh\left(\frac{\phi}{\sqrt{3}}\right)
\,, 
\end{eqnarray}
and it is a particular case of an Einstein-Maxwell-dilaton system. The solution of the corresponding  equations of motion are \cite{Cvetic:1999xp,Gubser:2000mm} 
\begin{eqnarray}
\label{GR_metric}
ds^{2}
=
\frac{r^2 }{L^2}
\left(\frac{Q}{r}+1\right)^{3/2}
\left(
-f(r)dt^{2}+d\vec x^{2}\right)
+
\frac{L^2
dr^{2}
}{
r^2 
f(r)
} \left(\frac{Q}{r}+1\right)^{-3/2}
,
\\ 
A_{t}=
\frac{\sqrt{3Q\tilde \mu}}{r+Q}
-
\frac{\sqrt{3Q\tilde \mu^{1/3}}}{L^{2/3}};
\qquad 
\phi=  \frac{\sqrt{3}}{2}
\ln\left(
1+\frac{Q}{r}
\right), \quad f=1-\frac{\tilde\mu L^{2}}{(r+Q)^{3}}.
\end{eqnarray}

\noindent 
In the above equations, $Q$ is related to the black hole charge, $\tilde\mu>0$ is related to the black hole mass and $L$ is the AdS radius. 
The black hole horizon is located at $r=(\tilde\mu L^2)^{1/3}-Q\equiv r_{h}$. 
Defining the new coordinate $\tilde r = r+Q$ so that $\tilde r_h = r_h+Q$, then
\begin{equation}
f = 1 - \frac{\tilde\mu L^2}{{\tilde r}^{3}} \,;\qquad
A_t = 
 -   \frac{\sqrt{3Q\tilde\mu^{1/3}}}{ L^{2/3}} \left[1- \frac{\tilde r_h}{\tilde r}
\right]\,,
\end{equation} 
which is similar to that of a Reissner-Nordstrom charged black hole with chemical potential given by 
\begin{equation}
\mu = 
   \frac{\sqrt{3Q\tilde\mu^{1/3}}}{ L^{2/3}} \,.
\end{equation} 
The Hawking temperature, using the $r$ coordinate defined in Eq. \eqref{GR_metric}, is then given by
\begin{eqnarray}
T
=
\frac 3{4\pi L^2}  \left(\tilde \mu L^2\right)^{1/6} \sqrt{r_h}
\end{eqnarray}
Then, the imaginary part of the admittance can be written from Eq. \eqref{general_adm_im} as a function of the temperature as
\begin{eqnarray}
\Im\chi(\omega)
=
\frac{3\alpha'}{2\omega T} \left(\tilde \mu L^2\right)^{-1/3}
=
\frac{9\alpha'Q}{2\omega T \mu^{2} L^2}
\,, 
\end{eqnarray}
while the real part from Eq. \eqref{real_adm} is
$\Re\chi \approx - {2\pi \alpha '}/{3r_b^3}\,.$ 
Note that the real part of the admittance is small compared with its
imaginary part since the former does not depend on the frequency $\omega$ which is small. 

The diffusion coefficient for this model, from  Eq.\eqref{Diffusion_Coef}, reads
\begin{eqnarray}
D=\frac{9Q\alpha'}{2\mu^{2}L^{2}}\,,
\end{eqnarray}
which is temperature independent and  decreasing for a growing chemical potential.

\subsection{Bosons}

The bosonic case is characterized by imposing the condition on the chemical potential $\mu < 0$. Then, 
for short times $t\ll \beta$, Eq. \eqref{bostpeq} gives 
\begin{eqnarray}\label{bostpeq2}
     s^{2}_{\rm Breg}(t)
     &\approx&
     \frac{2\alpha' }{g_{xx}(r_{h})}
     \text{Li}_2\left(e^{\beta\mu }\right)
     \frac{t^{2}}{\beta^{2}}
     =
     \frac{9\alpha'Q }{
     2\pi\mu^{2}L^{2} T}
     \text{Li}_2\left(e^{\beta\mu }\right)
     \frac{t^{2}}{
     \beta^{2}}
     \,, 
\end{eqnarray}
which decreases when $|\mu|$ increases. 

On the other side, for long times,  $t\gg\beta$, from Eq.\eqref{bostgran}, one finds 
\begin{eqnarray}
     s^{2}_{\rm Breg}(t)
     &\approx&\frac{4\alpha' }{g_{xx}(r_{h})}
     e^{\beta\mu}\log \left( \frac{t}{ \beta} \right)
     =
     \frac{9\alpha'Q}{
     \pi\mu^{2}L^{2}T} e^{\beta\mu}
     \log \left( \frac{t}{ \beta} \right)
     \,.
\end{eqnarray}
Since $\mu<0$ the diffusion also  drops down when $|\mu|$ grows.

\subsection{Fermions}

The fermionic case is described by $\mu>0$. So, for short times, from Eq.\eqref{fertpeq}, we obtain:
\begin{equation}
    s^2_{\rm Freg}(t)
    \approx
    \frac{\alpha' }{g_{xx}(r_{h})}\mu^{2}t^2
    =
    \frac{9\alpha'Q }{
    2\pi L^{2} T}t^2
    \,. 
\end{equation}
Note that if one keeps $Q$ fixed, there is no dependence on chemical potential for this result.  
On the other hand, for long times we can use the results in Eq. \eqref{fertgran} to obtain: 
\begin{equation}
 s^2_{\rm Freg}(t) \approx
 \frac{4\alpha' }{g_{xx}(r_{h})} \log \left(t\mu \right)
 = 
 \frac{9\alpha'Q}{
 \pi\mu^{2}L^{2}T}
 \log \left(t\mu \right),
\end{equation}
which diminishes for growing $\mu$.

%

\section{Conclusions}\label{sec:conc}

In this work we presented some results regarding the dynamics of a particle in a thermal bath at finite density, in the  linear regime, using holographic methods. For a general diagonal metric, we obtained expressions for linear response function, mean square displacement, correlations functions and diffusion coefficient, in terms of the metric elements. As we showed here, the mean square displacement differs for fermions and bosons. Also, we have verified the fluctuation-dissipation theorem for zero and non-zero chemical potentials for both statistics. As these results are presented in terms of the metric components, this is a quite general result.

Since we have a general relation between different metric elements and physical quantities as the admittance, the regularized mean square displacement and temperature, it is possible to choose particular backgrounds in order to produce systems with holographic duals that may have some desired physical behavior.

For all these results we see a dependence on the IR part of the metric in the direction of the electrical field on the brane, which acts as an external force for the system. This is quantified by the presence of $g_{xx}(r_{h})$ in the general expressions. 

For the admittance, we have found the hydrodynamic behavior $\omega^{-1}$ of its imaginary part, in accordance with the literature \cite{deBoer:2008gu, Tong:2012nf, Edalati:2012tc,Giataganas:2018ekx}. However, one finds different temperature behaviors with respect to different choices of the metrical components. Note also that the admittance does not depend on the statistics. 

For the regularized mean square displacement for bosons we found, in the late times limit, that the effect of turning on the chemical potential makes the diffusion process slower. The usual behavior $s^{2}_{\text{reg}}\sim t$ then becomes a Sinai-like diffusion, $s^{2}_{\text{reg}}\sim \log t$. On the other hand, for fermions, we always find $s^{2}_{\text{reg}}\sim \log t$ irrespective  the value of the chemical potential.

In the first application, we studied the Hyperscaling-Lifshitz background exploring different values for the parameters $z$ and $\theta$ for varied number of spatial dimensions $d$. It interesting to say that using this freedom of choice it was possible to obtain some interesting cases without violating the thermodynamic stability. For instance, the usual Einstein linear dependence of the temperature in the diffusion constant can be found, for a certain choice of parameters ($d=\theta$).    

For the Gubser-Rocha model,  Section  \ref{sec:GRM}, our second application, we found that the diffusion coefficient does not depend on the temperature and it is inversely proportional to the chemical potential. For bosons, the regularized mean square displacement decreases when the absolute value of chemical potential increases. Also, this quantity is inversely proportional to the temperature. As expected, the presence of a non-null chemical potential makes the diffusion processes becomes slower. The same behavior is found for the fermionic case, but now $s^{2}_{\text{reg}}\sim \log t$ and it is inversely proportional to the square of temperature. 

We presented here calculations allowing us to determine the diffusion coefficient, the admittance and the mean square distance for fermions which are relevant for systems with Fermi surfaces and Fermi liquids.
We hope that the discussion presented here might help finding new applications for holographic systems, in particular for fermionic ones. 


\section*{Acknowledgments}
We would like to acknowledge Luca Moriconi for discussions, Eduardo Capossoli for collaboration in the begining of this work and an anonymous referee for useful comments and suggestions. 
 N.G.C. is supported by  Conselho Nacional de Desenvolvimento Científico e Tecnológico (CNPq) and Coordenação de Aperfeiçoamento de Pessoal de Nível Superior (CAPES) under finance code 001. H.B.-F. and C.A.D.Z. are partially supported by Conselho Nacional de Desenvolvimento Cient\'{\i}fico e Tecnol\'{o}gico (CNPq) under the grants $\#$s 311079/2019-9 and  310703/2021-2, respectively.  C.A.D.Z. is also partially supported by Funda\c{c}\~{a}o Carlos Chagas Filho de Amparo \`{a} Pesquisa do Estado do Rio de Janeiro (Faperj) under Grant E-26/201{.}447/2021 (Programa Jovem Cientista do Nosso Estado).

\appendix

\section{Solution for the EoM}\label{sec:appA}

Here in this Appendix we are going to find the solution for the EoM, Eq.  \eqref{equation_motion}, following Refs. \cite{deBoer:2008gu,  Tong:2012nf, Edalati:2012tc, Giataganas:2018ekx}. First we will show how to rewrite the EoM into a convenient form using the Regge-Wheeler (tortoise) coordinates as a Schr\"{o}dinger-like equation in Appendix \ref{sec:appA1}. We then obtain the solution  near the horizon in Appendix \ref{sec:appA2}.  The solutions for the other regions are discussed in Section \ref{sec:Hydro}.

\subsection{Regge-Wheeler coordinate and Schr\"{o}dinger-like equation}\label{sec:appA1}

First, we assume that the metric elements
$a(r),b(r)$ and $c(r)$, defined in Eq.\eqref{eq:metriccomponents}, are regular functions at $r=r_{h}$  and 
the function $f(r)$, dubbed the \textit{horizon function}, has the following asymptotic values: 
\begin{equation}\label{eq:asympf}
    \lim_{r\to r_{h}} f(r) \sim f'(r_{h})(r-r_{h});
    \qquad \textrm{and}
    \qquad 
    \lim_{r\to \infty} f(r)=1.
\end{equation}

The equation of motion, Eq.  \eqref{equation_motion}. up to quadratic terms in the Nambu-Goto action reads
\begin{equation}
    \frac{\partial}{\partial r}\left(\frac{g_{xx}\sqrt{g_{tt}}}{\sqrt{g_{rr}}}\frac{\partial\delta x}{\partial r}\right)-\frac{g_{xx}\sqrt{g_{rr}}}{\sqrt{g_{tt}}}\frac{\partial^{2}\delta x}{\partial t^{2}}=0.
\end{equation}
Substituting the following Fourier decomposition
\begin{equation}
    \delta x(t,r)=h_{\omega}(r)e^{-i\omega t}
\end{equation}
into the above equation of motion, one finds
\begin{equation}\label{eq:eomr}
    \frac{\partial}{\partial r}\left(\frac{g_{xx}\sqrt{g_{tt}}}{\sqrt{g_{rr}}}\frac{\partial h_{\omega}}{\partial r}\right)+\omega^{2}\frac{g_{xx}\sqrt{g_{rr}}}{\sqrt{g_{tt}}}h_{\omega}=0.
\end{equation}
It is convenient to use the Regge-Wheeler coordinate. It naturally comes if we rewrite the line element as
\begin{equation}
    ds^{2}=g_{tt}\left(dt^{2}-\frac{g_{rr} dr^{2}}{g_{tt}}\right)+\cdots 
\end{equation}
Then, the Regge-Wheeler coordinate $r_{*}$ is defined by
\begin{equation}\label{eq:r*def}
    dr_{*}^{2}=\frac{g_{rr} dr^{2}}{g_{tt}} \Rightarrow dr_{*}=\frac{\sqrt{g_{rr}}}{\sqrt{g_{tt}}}dr.
\end{equation}
The boundary conditions in this new coordinate are $\lim_{r\to r_{h}}r_{*}(r)\to-\infty$ and \break $\lim_{r\to \infty}r_{*}(r)\to 0$. So, the equation of motion  \eqref{eq:eomr} can be cast in the form
\begin{align} 
    \frac{d^{2}h_{\omega}}{dr_{*}^{2}}+\frac{d\log(g_{xx})}{dr_{*}}\frac{dh_{\omega}}{dr_{*}}+\omega^{2}h_{\omega}&=0 \label{eq:eomr*}\,.
\end{align}
%

We rewrite the above equation as 
\begin{equation}\label{eq:eom}
    \frac{d^{2}h_{\omega}}{dr_{*}^{2}}+\alpha(r_{*})\frac{dh_{\omega}}{dr_{*}}+\beta(r_{*})h_{\omega}=0
\end{equation}
which can be transformed  into Hill's standard form \cite{Hill}
\begin{equation}\label{Hilleq}
    \frac{d^2\psi_{\omega}}{dr_{*}^{2}}+Q(r_{*})\psi_{\omega}=0, 
\end{equation}
by putting
\begin{align}
    h_{\omega}(r_{*})&=\exp\left[-\frac{A(r_{*})}{2}\right]\psi_{\omega},\label{def:psi}\\
    \alpha(r_{*})&=\frac{dA}{dr_{*}},\label{def:A}\\
    Q(r_{*})&=-\frac{1}{2}\frac{d\alpha}{dr_{*}}-\frac{1}{4}\alpha^{2}+\beta. \label{def:V}
\end{align}

Comparing Eq. \eqref{eq:eomr*} with \eqref{eq:eom} we have:
\begin{equation}
\alpha(r_{*})=\frac{d\log{g_{xx}}}{dr_{*}}\;\;\;\textrm{e}\;\;\;\beta(r_{*})=\omega^{2}.    
\end{equation}
Equation \eqref{def:A} gives
\begin{equation}
    \frac{d\log{g_{xx}}}{dr_{*}}=\frac{dA}{dr_{*}} \Rightarrow A(r_{*})=\log{g_{xx}}.
\end{equation}
Substituting the above result into Eq.  \eqref{def:psi} we get
\begin{eqnarray}
 h_{\omega}(r_{*})&=&\exp\left[-\frac{A(r_{*})}{2}\right]\psi_{\omega} \cr \cr 
 &=&\exp\left[-\frac{\log{g_{xx}}}{2}\right]\psi_{\omega}\cr\cr
 &=&\frac{\psi_{\omega}(r_{*})}{\sqrt{g_{xx}}}\label{eq:psifinal}
\end{eqnarray}
and $Q(r)$, Eq. \eqref{def:V}, is given by
\begin{eqnarray}
    Q(r)&=&-\frac{1}{2}\frac{d\alpha}{dr_{*}}-\frac{1}{4}\alpha^{2}+\beta \cr &=&-\frac{1}{2}\frac{dr}{dr_{*}}\frac{d}{d r}\left(\frac{dr}{dr_{*}}\frac{d\log{g_{xx}}}{dr}\right)
     -\frac{1}{4}\left(\frac{dr}{d r_{*}}\frac{d\log{g_{xx}}}{dr}\right)^{2}+\omega^{2}\nonumber \\
     &=&-\frac{1}{2}\frac{\sqrt{g_{tt}}}{\sqrt{g_{rr}}}\frac{d}{d r}\left(\frac{\sqrt{g_{tt}}}{\sqrt{g_{rr}}}\frac{d\log{g_{xx}}}{dr}\right)-\frac{1}{4}\frac{g_{tt}}{g_{rr}}\left(\frac{d\log{g_{xx}}}{dr}\right)^{2}+\omega^{2}. \label{eq:Q(r)}
\end{eqnarray}
%


Then, Hill's equation  \eqref{Hilleq}, can be written as a Schr\"{o}dinger-like equation:
\begin{equation}\label{eq:schrodinger}
    \frac{d^{2}\psi_{\omega}}{dr_{*}^{2}}+(\omega^{2}-V(r))\psi_{\omega}=0,
\end{equation}
where $\psi_{\omega}$ is defined as 
$\psi_{\omega}(r_{*})=\sqrt{g_{xx}}h_{\omega}(r_{*})$, and the potential is given by
\begin{equation}
    V(r)=\frac{1}{2}\frac{\sqrt{g_{tt}}}{\sqrt{g_{rr}}}\frac{\partial}{\partial r}\left(\frac{\sqrt{g_{tt}}}{\sqrt{g_{rr}}}\frac{d\log(g_{xx})}{dr}\right)+\frac{1}{4}\frac{g_{tt}}{g_{rr}}\left(\frac{d\log(g_{xx})}{dr}\right)^{2}.
\end{equation}
 In terms of the metric functions, Eq. \eqref{eq:metriccomponents}, this potential reads 
\begin{equation}\label{eq:potential}
    V(r)=\frac{f(r)}{2}\sqrt{\frac{a(r)}{b(r)}}\frac{d}{dr}\left(\sqrt{\frac{a(r)}{b(r)}}\frac{f(r)}{c(r)}\frac{d c(r)}{dr}\right)+\frac{1}{4}\frac{a(r)}{b(r)}\frac{f^{2}(r)}{c^{2}(r)}\left(\frac{d c(r)}{dr}\right)^{2}.
\end{equation}
Notice that $V(r_{h})=0$ as expected as $f(r_{h})=0$.  Furthermore it has units of squared energy. Now, one can obtain the solution of the Schr\"odinger-like equation \eqref{eq:schrodinger}, into two different regions: near by and far from the horizon.

\subsection{Region A: Deep IR}
\label{sec:appA2}

The deep IR regime is the near horizon region, defined by 
\begin{equation}
    r\sim r_{h},\;\;\; V(r)\ll\omega^{2}. 
\end{equation}
This happens because $V(r)$ is modulated by $f(r)$, and near the horizon this function becomes arbitrarily close to zero as can be seen in Eq. \eqref{eq:potential}.  

In this region, the Schr\"odinger-like equation  \eqref{eq:schrodinger} is written as
\begin{equation}\label{eq:psicompletehorizon}
    \frac{d^{2}\psi_{A\omega}}{dr_{*}^{2}}+\omega^{2}\psi_{A\omega}=0 \Rightarrow \psi_{A\omega}=A_{1}e^{-i\omega r_{*}}+A_{2}e^{i\omega r_{*}}.
\end{equation}
The tortoise coordinate $r_{*}$, Eq. \eqref{eq:r*def}, is given by
\begin{equation}\label{eq:r*horizon}
    \frac{d r_{*}}{d r}=\frac{\sqrt{g_{rr}}}{\sqrt{g_{tt}}}\approx\sqrt{\frac{b(r_{h})}{a(r_{h})}}\frac{1}{f'(r_{h})(r-r_{h})}\Rightarrow r_{*}\approx\frac{1}{f'(r_{h})}\sqrt{\frac{b(r_{h})}{a(r_{h})}}\log\left(\frac{r}{r_{h}}-1\right),
\end{equation}
where  Eq. \eqref{eq:asympf} was used. From Eq. \eqref{eq:psifinal} and considering only the ingoing solution ($A_{2} = 0$) in Eq. \eqref{eq:psicompletehorizon}, one finds
\begin{equation}
    h_{A\omega}(r)=\frac{A_{1}e^{-i\omega r_{*}}}{\sqrt{g_{xx}(r_{h})}}.
\end{equation}
Since one is interested in the long wavelength limit ($\omega\to 0$), using \eqref{eq:r*horizon}, the above equation reads
\begin{equation}
\label{hAp}
    h_{A\omega}(r)=\frac{A_{1}}{\sqrt{g_{xx}(r_{h})}}\left[1-i\frac{\omega}{f'(r_{h})}\sqrt{\frac{b(r_{h})}{a(r_{h})}}\log\left(\frac{r}{r_{h}}-1\right)\right].
\end{equation}
Note also that $Q(r)$, Eq. \eqref{def:V}, near the horizon becomes 
$Q(r)\approx \omega^{2}$, since  $r\approx r_{h}$ and using Eq. \eqref{eq:metriccomponents}, one gets 
\begin{equation}
\frac{\sqrt{g_{tt}}}{\sqrt{g_{rr}}}\approx \sqrt{\frac{a(r_{h})}{b(r_{h})}}\lim_{r\to r_{h}}f(r)=0.
\end{equation}

\section{Neumann boundary condition and Normalization}\label{sec:neumann}

\subsection{Neumann boundary condition}\label{AppB1}

The coefficient $B$ from Eq. \eqref{General_h_UV} can be obtained from a Neumann boundary condition at UV in the position $r=r_{b}$. At zero order in $\omega$, it is
simply 
\begin{eqnarray}
  B= 1\,. 
\end{eqnarray}
In general, keeping higher orders in $\omega$ this coefficient will be a pure phase $e^{i\omega\theta}$ for some $\theta$ real. This will happen because the general solution can always be written as
\begin{equation}
h^{UV}_{\omega}(r)
=  {A(\omega)} \left[g(r)
+  {B(\omega)}g^{*}(r)\right],
\end{equation}
that implies 
\begin{eqnarray}
\label{Pure_Phase}
        {B(\omega)}=-\left.\frac{\partial_{r} g(r)}{\partial_{r}g^{*}(r)}\right|_{r
  =r_{b}}
  =e^{i\omega\theta}. 
\end{eqnarray}
From a Neumann boundary condition near the horizon the coefficient $B$ can obtained 
\begin{equation}
    {B(\omega)}=-\left.\frac{\partial_{r}h^{in}(r)}{\partial_{r}h^{out}(r)}
  \right|_{r/r_{h}
  =1+\epsilon}
  =\exp\left\{-2i\frac{\omega}{f'(r_{h})}
  \sqrt{\frac{b(r_{h})}{
  a(r_{h})}}
  \log\left(\frac{1}{\epsilon}\right)\right\}\,,
\end{equation}
which is a pure phase in $\omega$. 


\subsection{Normalization}\label{AppB2}

Now we proceed to find the coefficient $A$ in Eq. \eqref{General_h_UV}. With this purpose, we write the Klein Gordon inner product, as 
\begin{equation}
\label{KGIP}
(f,g)
=-\frac{i}{2\pi\alpha'}
\int_{\Sigma}
dx\sqrt{|h|}
n^{\mu}
g_{xx}(f\partial_{\mu}g^{*}-\partial_{\mu}fg^{*}),
\end{equation}
where $h$ is the induced metric on the Cauchy surface $\Sigma$,  $n^{\mu}$ is the normal vector to this surface, $f$ and $g$ are solutions of the equations of motion.

As a Cauchy surface we take a constant time slice of our worldsheet. Therefore we have in the present situation
\begin{eqnarray}
   n^{\mu}
   =\left(\frac{1}{
   \sqrt{g_{tt}}},0\right) ,&& h=g_{rr}.
\end{eqnarray}      
Thus the Klein-Gordon inner product becomes
\begin{eqnarray}
\label{KGIP2}
(f,g)
=-\frac{i}{2\pi\alpha'}
\int_{\Sigma}
dx\sqrt{\frac{g_{rr}
}{g_{tt}}}
g_{xx}(f\partial_{t}g^{*}-\partial_{t }fg^{*}).
\end{eqnarray}

In order to obtain a proper  normalization, we demand  that $(X(t,r),X(t,r))=1$. Then, for $X(t,r)=e^{-i\omega t}h_{\omega}(r)$ we get
\begin{eqnarray}
    (X(t,r),X(t,r))
    &=&\frac{\omega}{\pi\alpha'}
    \int_{r_{h}}^{r_{b}}
    dr\sqrt{
    \frac{g_{rr}
    }{g_{tt}}}
    g_{xx}
    |h_{\omega}(r)|^{2}
    \cr
    &=&\frac{\omega}{\pi\alpha'}|  {A(\omega)}|^{2}
    \int_{r_{h}}^{r_{b}}
    dr\sqrt{
    \frac{b(r)
    }{
    a(r)}}
    \frac{g_{xx}(r)}{f(r)}
    |h_{\omega}(r)|^{2}
\end{eqnarray}

For regular functions $b(r)$ and $a(r)$ this integral will be dominated by the near horizon region because of the zero in $f(r)$. Therefore, we can approximate this integral by
\begin{eqnarray}
    (X(t,r),X(t,r))
    &\approx&\frac{\omega}{\pi\alpha'}|  {A(\omega)}|^{2}
    \sqrt{
    \frac{b(r_{h})
    }{
    a(r_{h})}}
    \frac{g_{xx}(r_{h})}{f'(r_{h})}
    \int_{r_{h}}^{r_{b}}
    \frac{dr}{r_{h}}
    \frac{
    |h_{\omega}(r)|^{2}}{
    \frac{r}{r_{h}}-1}
    \cr
    &\approx&\frac{\omega}{\pi\alpha'}|  {A(\omega)}|^{2}
    \sqrt{
    \frac{b(r_{h})
    }{
    a(r_{h})}}
    \frac{g_{xx}(r_{h})}{f'(r_{h})}
    \int_{r_{h}}^{r_{b}}
    \frac{dr}{r_{h}}
    \frac{
    1}{
    \frac{r}{r_{h}}-1},
\end{eqnarray}
where we used the near horizon expression for $h_{\omega}(r)$ Eq.\eqref{General_h_IR}, keeping only terms smaller than $O(\omega^{2})$. 
The final result is
\begin{eqnarray}
    (X(t,r),X(t,r))
    &\approx&\frac{\omega}{\pi\alpha'}|  {A(\omega)}|^{2}
    \sqrt{
    \frac{b(r_{h})
    }{
    a(r_{h})}}
    \frac{g_{xx}(r_{h})}{f'(r_{h})}
    \int_{r_{h}}^{r_{b}}
    \frac{dr}{r_{h}}
    \frac{
    1}{
    \frac{r}{r_{h}}-1}
    \cr
    &\sim& \frac{\omega}{\pi\alpha'}|  {A(\omega)}|^{2}
    \sqrt{
    \frac{b(r_{h})
    }{
    a(r_{h})}}
    \frac{g_{xx}(r_{h})}{f'(r_{h})}
   \log\left(\frac{1}{\epsilon}\right),
\end{eqnarray}
where the integral was regularized considering in the lower limit of integration \break $r_{h}\to r_{h}+\epsilon$. 
Imposing the normalization condition we obtain
\begin{eqnarray}\label{NormA}
      {A(\omega)}=\sqrt{\frac{\pi\alpha'f'(r_{h})}{4\omega g_{xx}(r_{h})\log\left(\frac{1}{\epsilon}\right)}\sqrt{\frac{a(r_{h})
    }{b(r_{h})
    }}}\,.
\end{eqnarray}

  {
\section{High frequencies suppression}\label{energy_scale}
In the text calculations we see integrals in the form 
\begin{eqnarray}
\label{baseint}
\int_{0}^{\infty} d\omega h(\omega)
\frac{g(\omega t)}{e^{\beta \omega} \pm 1}   
\end{eqnarray}
where $g(\omega t)$ is a periodic function such as $\cos(\omega t)$ or $\sin(\omega t)$, with maxima and minima for the arguments equal to $ n\pi$ with $n=0,1,2,...$ . 
Note that this kind of integral is dominated by the frequencies $\omega < T$, where $T=1/\beta$ is the temperature of the system.
Making the change of variables:
$$
\omega \to \frac{x}{t}
$$
the integral becomes
\begin{eqnarray}
\int_{0}^{\infty} \frac{dx}{t} h(x)
\frac{g(x)}{e^{k x} \pm 1}\, ,
\end{eqnarray}
with $k \equiv \beta/t $.
The function $\frac{g(x)}{e^{k x} \pm 1}$ works like a ``filter", each maximum (minimum) chooses one specific value of $x$ of the function $h(x)$. Those values are given by $x=n\pi$. For the $n$-th maximum (minimum) $P_n$ we have:
\begin{eqnarray}
    P_{n}
    =
    \frac{\pm 1}{e^{k n\pi} \pm 1}.
\end{eqnarray}
For $x_{n}=n\pi \gg 1/k$ we can approximate:
\begin{eqnarray}
    P_{n}
    \approx
    \pm e^{-kn\pi}. 
\end{eqnarray}
This means that for large values of $x_{n}$ the function $h(x)$ is exponentially suppressed. In terms of frequencies we can write that:
\begin{eqnarray}
x_{n}=\omega_{n} t \gg \frac{1}{k}=\frac{t}{\beta} \implies \omega_{n} \gg \frac{1}{\beta} = T
\end{eqnarray}
Thus, frequencies higher than the temperature are exponentially suppressed in the integral. Then, it is a good approximation to take frequencies smaller than the temperature $T$ in the function $h(\omega)$ in the integral.
In the case where the chemical potential is non-zero, the relevant integral is 
\begin{eqnarray}
\int_{0}^{\infty} d\omega h(\omega)
\frac{g(\omega t)}{e^{\beta (\omega-\mu)} \pm 1}    
\end{eqnarray}
after the change of variables
\begin{eqnarray}
 \omega \to y= t(\omega-\mu)   
\end{eqnarray}
this integral becomes
\begin{eqnarray}
\int_{-\mu t}^{\infty} \frac{dy}{t} h(y)
\frac{g(y)}{e^{k y} \pm 1}
\end{eqnarray}
where $k \equiv \beta/t$. Then, in an analogous way we conclude that for
\begin{eqnarray}
\label{omegamu}
    y_{n} =n\pi \gg
    \frac{1}{k}=\frac{t}{\beta}
    \cr
   \implies \omega_{n} -\mu \gg  T 
\end{eqnarray}
the function $h$ is exponentially suppressed.}

  {
Note that in the case of fermions we have $\mu \ge 0$ and then $\omega_{n} \ge 
\omega_{n} -\mu$ making that frequencies satisfying  
\begin{eqnarray}
    \omega_{n} \gg  T 
\end{eqnarray}
Become exponentially suppressed as well.}

  {
For bosons, in the limit of $|\mu| \ll \omega_{n}$ or $|\mu| \ll T$ we have that \eqref{omegamu} implies that frequencies $\omega_{n} \gg  T$ are suppressed too. On the other hand, in the case of $|\mu| \gg T$, $z = e^{-\beta \mu} \gg 1$ and we can write 
\begin{eqnarray}
\int_{0}^{\infty} d\omega h(\omega)
\frac{g(\omega t)}{e^{\beta (\omega-\mu)} - 1} 
=
\frac{1}{z}\int_{0}^{\infty} d\omega h(\omega)
\frac{g(\omega t)}{e^{\beta \omega} - \frac{1}{z}}\, ,
\end{eqnarray}
from this we can use the same argument carried out after Eq. \eqref{baseint} and conclude that frequencies 
\begin{eqnarray}
    \omega_{n} \gg  T\, , 
\end{eqnarray}
are essentially irrelevant for the integral.}


\begin{thebibliography}{ABC}	

\bibitem{deBoer:2008gu}
J.~de Boer, V.~E.~Hubeny, M.~Rangamani and M.~Shigemori,
JHEP \textbf{07}, 094 (2009)
doi:10.1088/1126-6708/2009/07/094
[arXiv:0812.5112 [hep-th]].


\bibitem{Tong:2012nf}
D.~Tong and K.~Wong,
``Fluctuation and Dissipation at a Quantum Critical Point,''
Phys. Rev. Lett. \textbf{110}, no.6, 061602 (2013)
[arXiv:1210.1580 [hep-th]].


\bibitem{Edalati:2012tc}
M.~Edalati, J.~F.~Pedraza and W.~Tangarife Garcia,
``Quantum Fluctuations in Holographic Theories with Hyperscaling Violation,''
Phys. Rev. D \textbf{87}, no.4, 046001 (2013)
[arXiv:1210.6993 [hep-th]].


\bibitem{Giataganas:2018ekx}
D.~Giataganas, D.~S.~Lee and C.~P.~Yeh,
``Quantum Fluctuation and Dissipation in Holographic Theories: A Unifying Study Scheme,''
JHEP \textbf{08}, 110 (2018)
[arXiv:1802.04983 [hep-th]].

\bibitem{Banerjee:2015vmo}
P.~Banerjee,
``Holographic Brownian motion at finite density,''
Phys. Rev. D \textbf{94}, no.12, 126008 (2016)
[arXiv:1512.05853 [hep-th]].




\bibitem{Caldeira:2020sot}
N.~G.~Caldeira, E.~Folco Capossoli, C.~A.~D.~Zarro and H.~Boschi-Filho,
``Fluctuation and dissipation from a deformed string/gauge duality model,''
Phys. Rev. D \textbf{102} (2020) no.8, 086005
doi:10.1103/PhysRevD.102.086005
[arXiv:2007.00160 [hep-th]].




\bibitem{Karch:2006pv}
A.~Karch, E.~Katz, D.~T.~Son and M.~A.~Stephanov,
``Linear confinement and AdS/QCD,''
Phys. Rev. D \textbf{74}, 015005 (2006)
[arXiv:hep-ph/0602229 [hep-ph]].

\bibitem{Andreev:2006ct}
O.~Andreev and V.~I.~Zakharov,
``Heavy-quark potentials and AdS/QCD,''
Phys. Rev. D \textbf{74}, 025023 (2006)
[arXiv:hep-ph/0604204 [hep-ph]].

\bibitem{Caldeira:2020rir}
N.~G.~Caldeira, E.~Folco Capossoli, C.~A.~D.~Zarro and H.~Boschi-Filho,
``Fluctuation and dissipation within a deformed holographic model with backreaction,''
Phys. Lett. B \textbf{815}, 136140 (2021)
doi:10.1016/j.physletb.2021.136140
[arXiv:2010.15293 [hep-th]].



\bibitem{Caldeira:2021izy}
N.~G.~Caldeira, E.~F.~Capossoli, C.~A.~D.~Zarro and H.~Boschi-Filho,
``Fermionic and bosonic fluctuation-dissipation theorem from a deformed AdS holographic model at finite temperature and chemical potential,''
Eur. Phys. J. C \textbf{82}, no.1, 16 (2022)
doi:10.1140/epjc/s10052-021-09963-3
[arXiv:2104.08397 [hep-th]].

\bibitem{sinai-like}
A. Bodrova, A. Chechkin, A. Cherstvy et al., ``Underdamped scaled Brownian motion: (non-) existence of the overdamped limit in anomalous diffusion'', Sci. Rep. \textbf{6} 30520 (2016)


\bibitem{Huijse:2011ef}
L.~Huijse, S.~Sachdev and B.~Swingle,
``Hidden Fermi surfaces in compressible states of gauge-gravity duality,''
Phys. Rev. B \textbf{85}, 035121 (2012)
doi:10.1103/PhysRevB.85.035121
[arXiv:1112.0573 [cond-mat.str-el]].


\bibitem{Gubser:2009qt}
S.~S.~Gubser and F.~D.~Rocha,
``Peculiar properties of a charged dilatonic black hole in $AdS_{5}$,''
Phys. Rev. D \textbf{81}, 046001 (2010)
[arXiv:0911.2898 [hep-th]].

\bibitem{Zubarev}
D. N. Zubarev, ``Nonequilibrium Statistical Thermodynamics,'' (Plenum Press, New York, 1974)
\bibitem{Markov:2009ue}
Y.~A.~Markov and M.~A.~Markova,
``On the fluctuation-dissipation theorem for soft fermionic excitations in a hot QCD plasma,''
Nucl. Phys. A \textbf{840}, 76-96 (2010)
[arXiv:0909.0377 [hep-ph]].

\bibitem{Zaanen:2015oix}
J.~Zaanen, Y.~W.~Sun, Y.~Liu and K.~Schalm, Holographic Duality in Condensed Matter Physics, Cambridge University Press, 2015.

\bibitem{novoselov}
K. Novoselov, A. Geim, S. Morozov et al. ``Two-dimensional gas of massless Dirac fermions in graphene''. Nature \textbf{438}, 197–200 (2005). https://doi.org/10.1038/nature04233

\bibitem{moriconi}
L. Moriconi, and D. Niemeyer. ``Graphene conductivity near the charge neutral point." Physical Review B \textbf{84}.19 (2011): 193401.



\bibitem{Dong:2012se}
X.~Dong, S.~Harrison, S.~Kachru, G.~Torroba and H.~Wang,
``Aspects of holography for theories with hyperscaling violation,''
JHEP \textbf{06}, 041 (2012)
doi:10.1007/JHEP06(2012)041
[arXiv:1201.1905 [hep-th]].


\bibitem{Hartnoll:2018xxg}
S.~A.~Hartnoll, A.~Lucas and S.~Sachdev, Holographic Quantum Matter,
MIT Press, 2018.
 

\bibitem{Gubser:2000mm}
S.~S.~Gubser and I.~Mitra,
``The Evolution of unstable black holes in anti-de Sitter space,''
JHEP \textbf{08}, 018 (2001)
doi:10.1088/1126-6708/2001/08/018
[arXiv:hep-th/0011127 [hep-th]].



\bibitem{Ogawa:2011bz}
N.~Ogawa, T.~Takayanagi and T.~Ugajin,
``Holographic Fermi Surfaces and Entanglement Entropy,''
JHEP \textbf{01}, 125 (2012)
doi:10.1007/JHEP01(2012)125
[arXiv:1111.1023 [hep-th]].

\bibitem{Kachru:2008yh}
S.~Kachru, X.~Liu and M.~Mulligan,
``Gravity duals of Lifshitz-like fixed points,''
Phys. Rev. D \textbf{78} (2008), 106005
doi:10.1103/PhysRevD.78.106005
[arXiv:0808.1725 [hep-th]].


\bibitem{Cvetic:1999xp}
M.~Cvetic, M.~J.~Duff, P.~Hoxha, J.~T.~Liu, H.~Lu, J.~X.~Lu, R.~Martinez-Acosta, C.~N.~Pope, H.~Sati and T.~A.~Tran,
``Embedding AdS black holes in ten-dimensions and eleven dimensions,'' Nucl. Phys. B \textbf{558} (1999), 96 doi:10.1016/S0550-3213(99)00419-8
[arXiv:hep-th/9903214 [hep-th]].




\bibitem{Hill}
W.~Magnus, S.~Wrinkler, Hill's Equation, Dover, 1979.
























\end{thebibliography}
\end{document}